\documentclass[twocolumn,showpacs,preprintnumbers,amsmath,amssymb]{revtex4}
\usepackage{CJK}
\usepackage{graphicx}
\usepackage{dcolumn}
\usepackage{bm}
\usepackage{color}
\numberwithin{equation}{section}
\begin{document}
\begin{CJK*}{GBK}{}
\preprint{APS/123-QED}

\title{The Smectic $A$-$C$ Phase Transition in Biaxial
Disordered Environments}
\author{Leiming Chen}
\address{College of Science, The China University of Mining and Technology, Xuzhou, Jiangsu, 200116, P. R. China}
\author{John Toner}
\affiliation{Department of Physics and Institute of Theoretical
Science, University of Oregon, Eugene, OR 97403}
\date{\today}
\begin{abstract}
We study the smectic $A$-$C$ phase transition in
biaxial
disordered environments, e.g. fully anisotropic aerogel. We find that both the $A$ and $C$ phases
belong to the universality class of  the `` XY Bragg glass '' , and therefore have
quasi-long-ranged translational smectic order. The phase transition itself belongs to a new universality class, which we study
using an $\epsilon=7/2-d$
expansion. We find a  stable fixed point, which implies a continuous transition, the critical exponents of which we calculate.
 \end{abstract} 
\maketitle
\end{CJK*}

\section{\label{Sec: Intro}Introduction}
The effect of quenched disorder on condensed matter systems
has been a widely studied topic for many years\cite{Harris,Geoff,Aharony}, both for practical reasons
(since  disorder is always present in real
systems) and fundamental ones. One of the fascinating results to emerge from the study of this problem is that  even {\it arbitrarily weak}  disorder can drastically affect the fundamental
properties of many systems. Among other effects, such weak disorder can destroy many types of long ranged order (e.g., ferromagnetic order in systems with quenched random fields \cite{RFferro}), and it can radically change the critical behavior of many phase transitions\cite{Aharony}.

Such effects have been found in, e.g., superconductors\cite{SC}, charge density waves\cite{CDW, CDWRT}, Josephson junction arrays\cite{JJunc}, superfluid helium in aerogel\cite{Helium}, and ferromagnetic superconductors\cite{SCferro}.

Some of the most novel and dramatic effects of quenched disorder are found in liquid crystals confined in random porous media\cite{LC,RT}. These intriguing systems exhibit a variety of exotic ``Bragg Glass" phases. They also undergo unique types of phase transitions\cite{CT}, one of which, the Smectic A to Smectic C (hereafter, AC) transition\cite{GP,ACexp}, is the subject of this paper.

In the high temperature phase ($A$ phase),  the nematic
director $\hat{n}$, which points along the axis of alignment of the constituent long molecules that make up the smectic material, and the normal to the smectic layers $\hat{N}$, are parallel.  In the low temperature phase ($C$ phase),
$\hat{n}$ and $\hat{N}$ tilt away from each other.

Here, we wish to study how this transition changes when the smectic is confined in a quenched, random, anisotropic environment. This can be achieved experimentally by confining the smectic within aerogel. By changing the preparation of the aerogel, it should be possible
to study the smectic $A$-$C$ transition in different anisotropic environments.

\begin{figure}
 \includegraphics[width=0.4\textwidth]{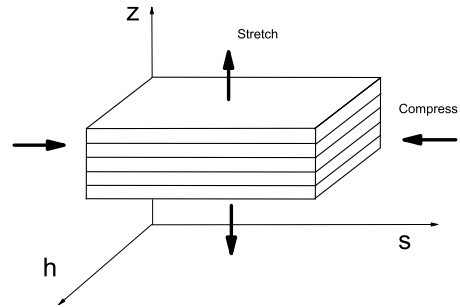}
 \caption{\label{fig: System}The geometry of the anisotropic disordered environment.}
\end{figure}

In previous work \cite{CT},  the nature of this transition was
studied for a liquid crystal confined in
a uniaxially stretched aerogel. In this paper we consider the
transition in a biaxial disordered environment, in which
the aerogel is first stretched along one axis, then compressed in an arbitrary direction perpendicular to the stretch\cite{brittle}. As illustrated in Fig. \ref{fig: System}, we refer the stretched direction as the $z$
axis, the compressed direction as the $h$
axis, and the direction perpendicular to both $\hat{z}$ and $\hat{h}$ as
the $s$ axis. We denote the entire  plane perpendicular to $\hat{z}$  (i.e., the $s-h$ plane)  as $\perp$.

We find that, in this geometry, the AC transition is a continuous transition belonging to a new universality class. Unusually, the upper critical dimension, below which the transition exhibits non-mean-field critical behavior, is $d_{UC}=7/2$. We have studied this phase transition using an $\epsilon={7\over
2}-d$ expansion, and find that there exists a
stable fixed point, which implies a second-order phase transition with {\it
universal} critical exponents. Furthermore, we find anisotropic scaling at the transition.

One of these universal exponents is the usual order parameter exponent $\beta$, defined via the temperature dependence of the tilt angle $\theta$ between the local molecular axis (i.e., the nematic
director $\hat{n}$), and the normal to the smectic layers $\hat{N}$.
As $T\to T_{AC}^-$, this angle, which is the magnitude of the order parameter for this transition, obeys the universal scaling law:
\begin{eqnarray}
\theta \propto (T_{AC}-T)^{\beta}
\end{eqnarray}
with
\begin{eqnarray}
 \beta={1\over 2}-{2\over 9}\epsilon+O(\epsilon^2).
 \label{betaeps}
\end{eqnarray}

In the high temperature phase ($A$ phase), both the nematic
directors $\hat{n}$ and the layer normals $\hat{N}$ lie, on
average along $\hat{z}$. In the low temperature phase ($C$ phase),
$\hat{n}$ and $\hat{N}$ tilt simultaneously away from $\hat{z}$ in opposite directions
in the $s-z$ plane.
Near $T_{AC}$, the angles $\theta_N(T)$ between $\hat{N}$ and $\hat{z}$
and $\theta_n(T)$ between $\hat{n}$ and $\hat{z}$ are not independent, but rather have the same
power-law temperature dependence as the total angle $\theta(T)\equiv\theta_N(T)-\theta_n(T)$ between $\hat{N}$ and $\hat{n}$. The ratio of $\theta_N(T)$ to $\theta_n(T)$ is a non-universal (negative) constant.

This geometrical constraint forbids azimuthal rotations of $\hat{n}$ and $\hat{N}$ in the $C$ phase. As a result,  there are no additional ``Goldstone modes" associated with the breaking of azimuthal symmetry in this anisotropic situation, in contrast to the isotropic case\cite{GP, CT}.
Indeed,  we find that both the $A$ and $C$ phases belong to the universality class of the random field $XY$ model (RFXY)\cite{Fisher, CT, Karl}.

The correlation length of the transition is defined as the length beyond which the two point real space correlations
of the order parameter $\vec{c}$ become small. Here $\vec{c}$ is the projection of $\hat{N}$ along $\hat{s}$; its magnitude is therefore
proportional to the tilt angle $\theta_N$. As $T\to T_{AC}$, the correlation lengths  $\xi_{s,h,z}$ along $\hat{s}$,
$\hat{h}$, and $\hat{z}$ become very different:  they diverge as power laws with different exponents:
\begin{eqnarray}
 \xi_{s,h,z}\propto |T-T_{AC}|^{\nu_{s,h,z}}.
 \label{nudef}
\end{eqnarray}
To  first order in $\epsilon$, we find
\begin{eqnarray}
 \nu_s&=&{1\over 2}+{7\over 27}\epsilon+O(\epsilon^2),\\
 \nu_z&=&1+{7\over 27}\epsilon+O(\epsilon^2),\\
 \nu_h&=&1+{10\over 27}\epsilon+O(\epsilon^2).
 \label{nu_eps}
\end{eqnarray}

The critical exponents most closely analogous to the exponent $\eta$ in conventional phase transitions\cite{Ma's book, eta} are those associated with ``anomalous elasticity" right at the critical point. Anomalous elasticity means that the elastic moduli characterizing the smectic are
not, as in conventional materials, constants,  but rather become dependent on the length scale $\lambda$ or wavevector $\vec{q}$ on which they are measured. More
specifically, the bend modulus $K$ that determines the free energy cost of bending the smectic layers diverges as $\vec{q}\to \vec{0}$
according to scaling laws
\begin{eqnarray}
 K(\vec{q})=q_s^{-\eta_K}f_K(q_h/q_s^{\zeta_h}, q_z/q_s^{\zeta_z});
\end{eqnarray}
while the compression modulus $B$ characterizing the resistance of the smectic layers to being squeezed closer together vanishes according to
\begin{eqnarray}
 B(\vec{q})=q_s^{\eta_B}f_B(q_h/q_s^{\zeta_h}, q_z/q_s^{\zeta_z}).
\end{eqnarray}
The anisotropy exponents $\zeta_h$ and $\zeta_z$ are defined as $\zeta_h\equiv2-\eta_K/2$ and
$\zeta_z\equiv2-(\eta_B+\eta_K)/2$. Our $\epsilon$-expansion results for $\zeta_{s,z}$ quoted below show that neither of these exponents equals one. This implies that the system exhibits anisotropic scaling. In fact, since $\zeta_h\ne\zeta_z$, there is completely anisotropic scaling between all three directions: soft ($s$), hard ($h$), and along the stretch axis ($z$).

The strength of the quenched disorder - specifically, of the random torques and random compression exerted by the
aerogel on the smectic layers - is characterized by the disorder invariance $\Delta_{t,c}$
where the subscripts $t,c$ respectively denote ``torque" and ``compression". This  also becomes wavevector dependent right at the AC transition, with an anisotropic scaling form:
\begin{eqnarray}
 \Delta_{t,c}(\vec{q})=q_s^{-\eta_{t,c}}f_{t,c}(q_h/q_s^{\zeta_h}, q_z/q_s^{\zeta_z}).
\end{eqnarray}
To leading order in $\epsilon={7\over2}-d$, the anomalous exponents $\eta_K$, $\eta_B$, and
$\eta_{t}$ are given  by
\begin{eqnarray}
 \eta_K&=&{16\over 27}\epsilon+O(\epsilon^2)\approx 0.3, \label{epsK}\\
 \eta_B&=&{4\over 9}\epsilon+O(\epsilon^2)\approx 0.22,\label{epsB}\\
 \eta_t&=&{4\over 27}\epsilon+O(\epsilon^2)\approx 0.07,\label{epsDt}\\
 \eta_c&=&2-{8\over 9}\epsilon+O(\epsilon^2)\approx 1.55\label{epsDc}.
\end{eqnarray}
The anisotropy exponents are:
\begin{eqnarray}
 \zeta_z&=&2-{(\eta_B+\eta_K)\over 2}=2-{14\over 27}\epsilon+O(\epsilon^2)\nonumber \\&\approx& 1.74~,
 \label{epszetaz}
 \\
 \zeta_h&=&2-{\eta_K\over 2}=2-{8\over 27}\epsilon+O(\epsilon^2)\nonumber \\
 &\approx& 1.85~.
\label{epszetah}
\end{eqnarray}

Finally, we find that the specific heat exponent $\alpha$ is given by
\begin{eqnarray}
 \alpha={2\over 9}\epsilon+O(\epsilon^2).
 \label{alphaeps}
\end{eqnarray}

Since $\epsilon=1/2$ is quite small in the physical dimension $d=3$, we expect that our predictions for $\alpha$, $\beta$,
and $\eta_{B,K,t}$ are quantitatively accurate.

In addition, there are several {\it exact} scaling relations between these critical exponents:
\begin{eqnarray}
\beta &=&{\nu_s\over 4}\left[4d-10+(5-d)\eta_K-\eta_B-2\eta_t\right],\label{ExactBeta}\\
\alpha&=&2-{\nu_s\over 2}\left[4d-6+(3-d)\eta_K-2\eta_t-\eta_B\right],\nonumber\\\label{ExactAlpha}\\
\eta_c&=&2+\eta_t-\eta_B-\eta_K,\label{ExactEta_c}\\
\nu_z&=&\zeta_z\nu_s,\label{ExactNu_z}\\
\nu_h&=&\zeta_h\nu_s.\label{ExactNu_h}
\end{eqnarray}
As in the $AC$ transition in uniaxially disordered environments \cite{CT}, $\alpha$ does {\it not} obey the usual anisotropic hyperscaling
relation $\alpha= 2-\nu_s-(d-2)\nu_h-\nu_z$; this is due to the
strongly relevant disorder.

The behavior of the correlation functions near the $AC$ transition is more complicated than in conventional phase transitions; in particular, it cannot be summarized by a single correlation length exponent $\nu$. This is in part due to the fact, mentioned above, that the transition
exhibits anisotropic scaling. It is also due to the fact that the correlations within the A and C phases themselves are quite non-trivial.

Since both the $A$ and $C$ phases belong to the $RFXY$ universality class, we can use the results of reference \cite{Karl} to obtain the x-ray scattering intensity (which is a probe of the smectic translational order) far away from the critical point. The most surprising feature of this is that neither phase shows true long-ranged translational order; rather, they exhibit {\it quasi}-long-ranged translational order, but with a {\it universal} exponent\cite{clean smectic, Caille}. The experimental signature of this quasi-long-ranged order is a power-law, rather than delta function, singularity of scattering as the Bragg peak is approached:
\begin{eqnarray}
I_n\left(\delta\vec{q}\right)\propto [(\delta q_s)^2+\alpha_h
(\delta q_h)^2+\alpha_z(\delta q_z)^2] ^{{-3+.55n^2}\over 2},
\label{Intensity}
\end{eqnarray}
where $n$ is the index of the Bragg peak, and $\delta\vec{q}=\vec{q}-\vec{G}_n$ with
$\vec{G}_n$ the reciprocal lattice vector. In the $A$ phase $\vec{G}_n={2n\pi\over a} \hat{z}$
with $a$ the layer spacing constant. In the $C$ phase $\vec{G}_n={2n\pi\over a}(\cos{\theta_N}\hat{z}\pm\sin{\theta_N}\hat{s})$, where ``$\pm$''
reflects that the tilting of the layer normals breaks two-fold symmetry, and ``$+$'' (``$-$'')
is taken if the tilting is along $\hat{s}$ ($-\hat{s}$). $\alpha_{h,z}$ are non-universal constants of
order 1. The exponent ${-3+.55n^2}\over 2$ implies that $I_n(\delta\vec{q})$
diverges as a power law only for the first two Bragg peaks .



In contrast, right at the AC critical point, the system only has short-ranged translational
order; as a result, the Bragg peaks are broad.

\begin{figure}
 \includegraphics[width=0.4\textwidth]{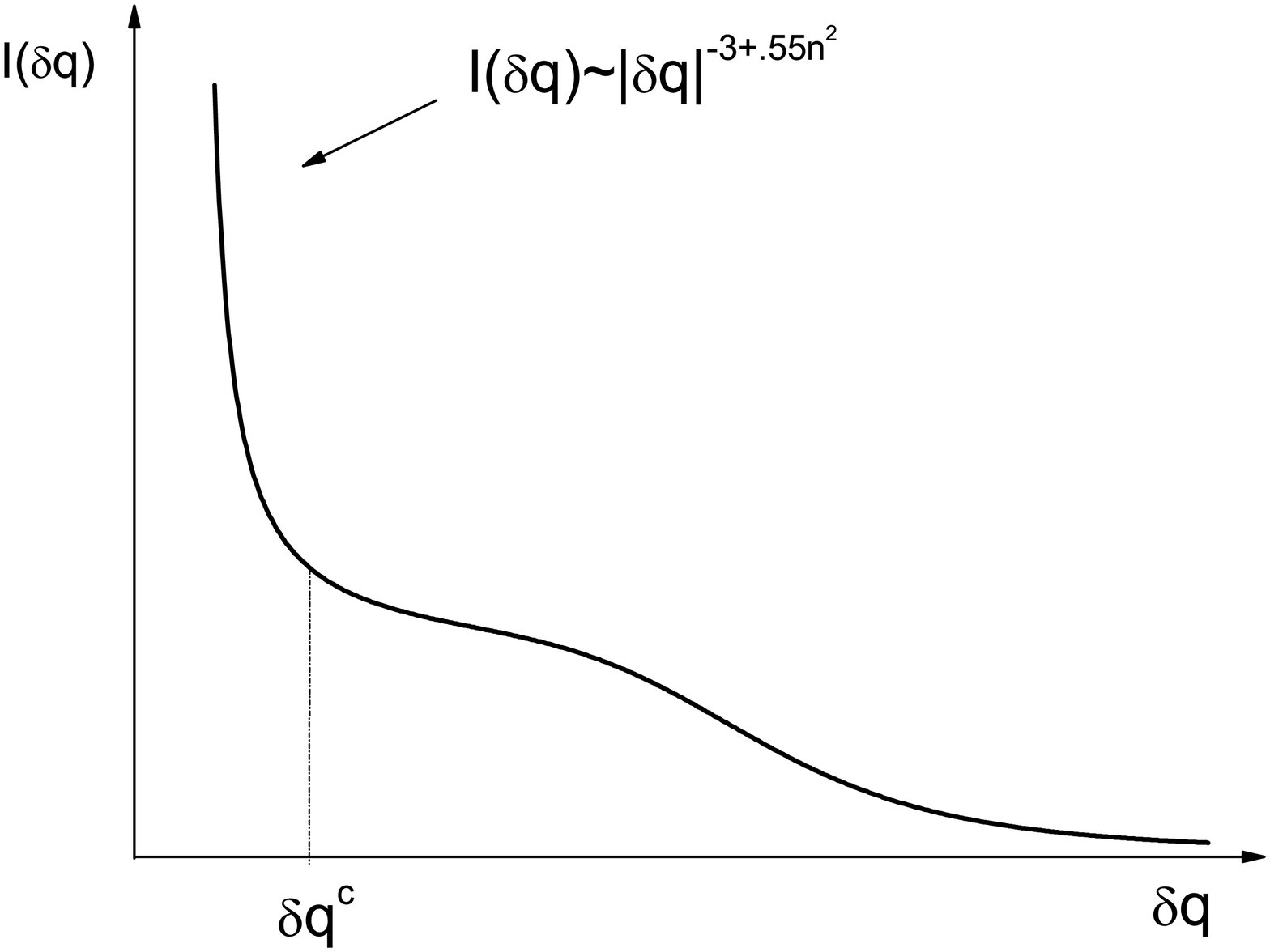}
 \caption{\label{fig: Xray}The X-ray scattering intensity as a function of $\delta q$,
 the deviation from the center of the first Bragg peak, near the critical point.}
\end{figure}

The X-ray scattering pattern near, but not at, the critical point interpolates between these two limits, as illustrated in Fig. \ref{fig: Xray}.
Sufficiently far away from the center of the Bragg peak, the line shape of
$I(\delta \vec{q})$ versus $\delta \vec{q}$ is qualitatively Lorenzian squared.
Close to the center of the Bragg peak, the line shape diverges as the
power law given by Eq. (\ref{Intensity}). The crossover between these two
distinct patterns - or, equivalently, the width of the spike on the top of the line shape, is
isotropic in scaling and strongly temperature dependent,
\begin{eqnarray}
 \delta q^{c}\propto \exp\left(-A|t|^{-\Omega}\right),
 \label{Omegadef}
\end{eqnarray}
where $A$ is non-universal and of order 1, and $t$ is the reduced temperature $t \equiv {T-T_{AC}\over T_{AC}}$.
$\Omega$ is universal,  and we can bound it above and below:
\begin{eqnarray}
 .25<\Omega<.65~~.
\end{eqnarray}
The width of the quasi sharp spike decreases extremely rapidly as $T\to T_{AC}$, and
vanishes completely right at $T=T_{AC}$.
Defining a correlation length $\xi$ for the AC transition via $\xi\equiv q_c^{-1}$, we see that $\xi$ diverges like a stretched exponential as the AC transition is approached;  hence, the usual correlation length exponent $\nu = \infty$, in this sense.

Another detailed experimental probe of our predictions is light scattering. Consider first an experiment in which
the magnitude $q$ of the wavenumber $q$ is varied for fixed, generic {\it direction} $\hat{q}$ of the scattering wavevector $\vec{q}$.  By ``generic direction $\hat{q}$ ", we mean a direction of $\hat{q}$ for which  no component of  $\hat{q}$ is $\ll 1$. We find that in such an experiment, in both the A and C phases, the light scattering intensity $I(\vec{q})$ plotted as a function of of $q$ shows three distinct scaling regimes:

\begin{eqnarray}
I(\vec{q})\propto\left\{
 \begin{array}{ll}
q^{-{\eta_c\over\zeta_h}}~~~~~~~~,~q\gg\xi_h^{-1}~,\\
\mbox{constant} ~~~~,~q^F(T) \ll q\ll\xi_h^{-1}\\
q^{-1}~~~~~~~~~~,~~ q \ll q^F(T)~,\\ \end{array}
 \right.
 \label{Iqgen}
\end{eqnarray}
where $q^F(T)$ vanishes as $T\rightarrow T_{AC}$ according
\begin{eqnarray}
q^F(\hat{q})\propto |T-T_{AC}|^{\phi}~,
\label{q_c}
\end{eqnarray}
where the universal exponent $\phi$ is given by
\begin{eqnarray}
\phi=\nu_s\left(1+\eta_c+{\eta_B-\eta_K\over 2}\right)={3\over 2}+{8\over 27}\epsilon+O(\epsilon^2)~.\nonumber\\
\label{phi}
\end{eqnarray}
The first equality in  (\ref{phi}) is exact, while the second follows from the known $\epsilon$-expansions (\ref{epsK}),(\ref{epsB}), and (\ref{epsDc}),  for $\eta_K$, $\eta_B$, and $\eta_c$.

\begin{figure}
 \includegraphics[width=0.4\textwidth]{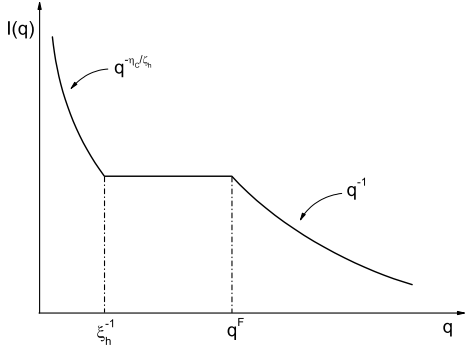}
 \caption{\label{fig: lightgen}Plot of $I(\vec{q})$, as given by equation (\ref{Iqgen}), versus the magnitude $q$ of $\vec{q}$ for a fixed ``generic'' direction $\hat{q}$ of $\vec{q}$, where ``generic direction'' is defined in the text.}
\end{figure}

This prediction for $I(\vec{q})$ as a function of $q$ for generic directions $\hat{q}$ of $\vec{q}$  is plotted in Fig. \ref{fig: lightgen}.

\begin{figure}
 \includegraphics[width=0.4\textwidth]{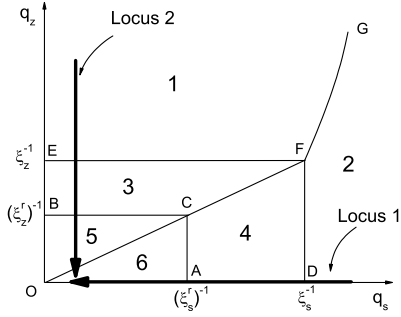}
 \caption{\label{fig: lightsz}The six regions of different forms for the light scattering in the $q_h=0$ plane.The boundaries between the different regions are given by equations (\ref{FG}-\ref{AC}), while the light scattering within each region is given by equation (\ref{Iqh=0}).}
\end{figure}

More information can be obtained by restricting the scattering to the subspace of $\vec{q}$ with $q_h=0$. In this case, we find that the light scattering exhibits  different behaviors  in the six different regions of the
 $q_h$-$q_z$, $q_h=0$ plane illustrated in Fig. \ref{fig: lightsz}.

In this figure, the various boundaries between regions are given by:
\begin{eqnarray}
FG:&&q_z\xi_z^N =
\left(q_{\perp}\xi_s^N\right)^{\zeta_z},\label{FG}\\
EF:&&q_z =\xi_z^{-1},\label{EF}\\
DF:&&q_s =\xi_s^{-1},\label{DF}\\
OF:&&q_z = {\xi_s\over\xi_z}q_s,\label{OF}\\
BC:&&q_z = \left(\xi_z^r\right)^{-1},\label{BC}\\
AC:&&q_s = \left(\xi_s^r\right)^{-1}.\label{AC}
\end{eqnarray}
and the light scattering in each region obeys:
\begin{widetext}
\begin{eqnarray}
 I(\vec{q})\propto\left\{
 \begin{array}{ll}
 q_s^2\left(q_z\xi_z^N\right)^{{2\eta_K-\eta_t-6\over\zeta_z}}, &\mbox{region 1}\\
q_s^{2\eta_K-\eta_t-4} \left(\xi_s^N\right)^{2\eta_K-\eta_t-6}, &\mbox{region 2}\\
q_s^2 \left(\xi_s^N\over\xi_s\right)^{{2\eta_K-\eta_t-6}}\left(q_z\xi_z\right)^{-2}, &\mbox{region 3}\\
 \left(\xi_s^N\over\xi_s\right)^{2\eta_K-\eta_t-6}\left(\xi_s\right)^{-2}, &\mbox{region 4}\\
q_s^2 \left(\xi_s^N\over\xi_s\right)^{1-{\eta_K\over 2}-\eta_B}\left(q_z\xi_z^N\right)^{-3}, &\mbox{region 5}\\
 \left(\xi_s^N\over\xi_s\right)^{\eta_K+{1\over 2}\eta_B-2}\left(\xi_s^N\right)^{-3}q_s^{-1},&\mbox{region 6}
 \end{array}
 \right.
 \label{Iqh=0}
\end{eqnarray}
\end{widetext}

The two new characteristic lengths $\xi_z^r$ and $\xi_s^r$  are given by:
\begin{eqnarray}
\xi_s^r&=&\xi_s^N\left(\xi_s\over\xi_s^N\right)^{\Gamma_s},\label{Gamma_s}\\
\xi_z^r&=&\xi_z^N\left(\xi_s\over\xi_s^N\right)^{\Gamma_z},
 \end{eqnarray}
with
\begin{eqnarray}
 \Gamma_s&=&2-\eta_K+\eta_t+{1\over 2}\eta_B,\\
 \Gamma_z&=&3-{3\over 2}\eta_K+\eta_t.
\end{eqnarray}

We also find that the transition temperature $T_{AC}$ depends on the anisotropy of the disorder.
Compared to the bulk transition temperature it is lowered by the
stretch of the aerogel but
raised by the compression. Hence, there is no a priori way to predict whether the net effect of the anisotropy is to raise or lower the AC transition temperature.

In this problem there are two ways continuing the model from 3 dimension to higher dimensions.
The $\epsilon=7/2-d$ expansion uses the hard analytical continuation, in which we take the $h$-space to be
$d-2$ dimensional, while both the $z$- and $s$-spaces are kept one dimensional. It is complemented by the
$\tilde{\epsilon}=4-d$ expansion, which uses the soft analytical continuation, where both $h$- and
$z$-space are one dimensional, and $s$-space is $d-2$ dimensional. The
$\tilde{\epsilon}$ expansion also leads to a stable non-Gaussian fixed point, which
again implies a second-order phase transition. However, the numerical
estimate of the exponents are not in good agreement with the $\epsilon$ expansion in $d=3$.
The technical details are given in Appendix \ref{Sec: SoftContinuation}.

The remainder of this paper is organized as follows. In
Sec. \ref{Sec: Model}, we derive our model for the $AC$ transition in biaxial
disordered environments. In Sec. \ref{Sec: RG}, we
derive the renormalization-group recursion relations for the
transition. In Secs. \ref{Sec: Critical Behavior} and  \ref{Sec: Anomalous Elasticity}, we calculate the thermodynamic critical
exponents, and  wavevector
dependences of the elastic constants and disorder variances, respectively.
In Sec. \ref{Sec: scat}, we determine the experimentally observable correlation functions
and make predictions for  X-ray and visible light scattering. In Sec. \ref{Sec: Stability}, we
show that the transition is stable against dislocations in the smectic layers.

\section{\label{Sec: Model}Model}
In keeping with earlier work on the AC transition in clean\cite{GP} and uniaxially disordered\cite{CT} systems, we assume that the local layer displacement $u(\vec{r})$ and the local nematic director $\hat{n}(\vec{r})$
are the only important fluctuating quantities in this problem \cite{deGennes}. The complete Hamiltonian, like Gaul, can be divided into
three parts. The first part is obtained by a slight modification of
the Hamiltonian for the smectic A phase in isotropic disordered environments\cite{RT,Karl,CT}.
It reads
\begin{eqnarray}
H_1 &\equiv& \int d^3r  \left[{ K \over 2}(\nabla^2_{\perp}u)^2 + {B
\over  2}(\partial_zu)^2 - {g \over 2}(\partial_zu)
|\vec{\nabla}_{\perp}u|^2 \right.\nonumber\\&+&\left. {w \over 8}
\left|\vec{\nabla}_{\perp}u \right|^4
+\vec{h}\left(\vec{r}\right)\cdot\vec{\nabla}u +
V_p(u-\phi(\vec{r}))\right] \label{H1}.
\end{eqnarray}
Since in our problem the liquid crystal is in an anisotropic environment, the Hamiltonian Eq. (\ref{H1})
need not be invariant under rigid rotations of the liquid crystal. Therefore, the values of the coefficients $B$, $g$ and $w$ are not
required to satisfy the constraint $B=g=w$, which they do in the isotropic case\cite{GP,deGennes}.
The $\perp$ and $z$ components of the random field $\vec{h}(\vec{r})$ in Eq. (\ref{H1}) respectively incorporate the random torques and random compressions exerted on the smectic by the aerogel. We take $\vec{h}(\vec{r})$ to be Gaussian, zero mean, and characterized by anisotropic short-ranged correlations:
\begin{eqnarray}
 \overline{h_i(\vec{r})h_j(\vec{r}\,')}=\left(\Delta_t\delta^{s}_{ij}+\Delta_h\delta^{h}_{ij}+\Delta_c\delta_{ij}^z\right)\delta^d(\vec{r}-\vec{r}\,')~, \nonumber\\
 \label{RFcorr}
\end{eqnarray}
where $\delta^{h}_{ij}=1$ for $i=j$, and both $i$ and $j$ taking values in one of the $d-2$ hard directions, and zero otherwise;  $\delta^{s}_{ij}=1$ for $i=j=s$, and zero otherwise; and $\delta^{z}_{ij}=1$ for $i=j=z$, and zero otherwise.
The function $V(u-\phi(\vec{r}))$ is a periodic function with period $a$, the smectic
layer spacing. The field $\phi(\vec{r})$ is a random length uniformly distributed between 0 and $a$ with
short-ranged correlations. Thus, $V(u-\phi(\vec{r}))$ represents the random pinning of the
smectic layers.

The second part of our Hamiltonian describes the coupling between layer normals $\hat{N}$ and directors $\hat{n}$.
The simplest such coupling is given by \cite{deGennes}
\begin{eqnarray}
H_2 &\equiv& \int d^dr
\ \ P|\hat{N}-\hat{n}|^2.
\label{Nn}
\end{eqnarray}

This part of the
Hamiltonian is responsible for the AC transition. To see this, simply note that when $P > 0$,
the lowest energy state has $\hat{N}\parallel\hat{n}$; this corresponds to the A phase. In contrast, when $P<0$, the energy can be lowered by making
$\hat{N}$ and $\hat{n}$ point in different directions; this corresponds to the C phase. Thus, the point $P=0$ corresponds, in mean field theory, and in the absence of other couplings between
$\hat{N}$ and $\hat{n}$, to the critical point for the pure smectic
$AC$ transition. If the coupling $P$ is assumed to be a monotonically decreasing
function of the temperature $T$, the A to C transition will occur as the temperature is lowered through the temperature at which $P=0$ ( again, in mean field theory and the absence of other couplings between $\hat{N}$ and $\hat{n}$).

Since the layer normal $\hat{N}$ is related to the smectic layer displacement $u$ by the geometric relation
\begin{eqnarray}
 \hat{N}={\hat{z}-\vec{\nabla}u\over |\hat{z}-\vec{\nabla}u|} ,
 \label{Geometric}
\end{eqnarray}
we can rewrite
Eq. (\ref{Nn}) as a coupling between $u$ and
 $\hat{n}$;
to lowest non-trivial order in $u$,  this reads:
\begin{eqnarray}
H_2 &\equiv& \int d^3r
\ \ P|\vec{\nabla}_{\perp}u+{\vec{\delta n}_{\perp}}|^2 .
\label{H2}
\end{eqnarray}


Since the aerogel is stretched and compressed, the disorder is no longer isotropic. Instead,
on average it picks out a preferred orientation for both the layer normals and the directors. Therefore,
in general the Hamiltonian should also include
\begin{eqnarray}
H_3 &\equiv& \int d^3r \left[M|\vec{\nabla}_{\perp}u|^2
+Q|{\vec{\delta n}_{\perp}}|^2\right.\nonumber\\&&+
\left.(M'-M)(\partial_hu)^2 + (Q'-Q)(\delta n_h)^2 \right] ,~~~~ \label{H3}
\end{eqnarray}
which is not rotation invariant and hence forbidden in the isotropic case.
It is expected that the stretching of the aerogel tends to hold both $\hat{N}$
and $\hat{n}$ along $\hat{z}$, while the compression tends to hold
both $\hat{N}$ and $\hat{n}$ within the $zs$ plane. Therefore, the coefficients
in Eq. (\ref{H3})obey the bounds
\begin{eqnarray}
M'>M>0~~~~,~~~~~~~~~~~ Q'>Q>0~~~~ .~~~~ \label{param constraint}
\end{eqnarray}
The values of
$M$ and $Q$  reflect how much the aerogel is stretched; those of $M'-M$ and $Q'-Q$ likewise reflect the degree of compression.

After a linear change of variables
\begin{eqnarray}
\delta n'_s&=&\delta n_s+R\partial_su~,\nonumber\\
\delta n'_h&=&\delta n_h +R'\partial_hu~,
\label{massive}
\end{eqnarray}
the sum of $H_2$ and $H_3$ becomes
\begin{eqnarray}
H_s&=& \int d^3r \left[{D(T)\over
2}(\partial_su)^2+(P+Q)(\delta n'_s)^2\right.\nonumber\\
&+&\left.{\gamma\over 2}(\partial_hu)^2+(P+Q')(\delta n'_h)^2\right] , \label{Hs}
\end{eqnarray}
where $R\equiv P/(Q+P)$, $R'\equiv P/(Q'+P)$, $D(T) \equiv 2M+2QP/(Q+P)$, $\gamma \equiv
2M'+2Q'P/(Q'+P)$. It is easy to verify that under the constraints (\ref{param constraint}) $D(T)<\gamma<P+Q<P+Q'$.
Therefore, if we start in the high temperature ($A$) phase and lower the temperature,
$D(T)$ is the first among the coefficients in Eq. (\ref{Hs})
to become negative.
As we will see in a moment, the critical point (ignoring fluctuations) is given by $D(T_{AC})=0$, which leads to
$P(T_{AC})= -MQ/(Q+M)$.
Therefore, stretching the aerogel lowers the critical temperature. Stretching aerogel suppresses the transition. Since $P_{AC}$ has no dependence on
$M'$ and $Q'$, $T_{AC}$ is not affected by the compression, in this mean field approximation in which fluctuations are neglected.

Now we look for the ground state of the system, averaged over many realizations of the disorder.
Since the disorder pieces in Eq. (\ref{H1}) are completely random and isotropic,
they have no averaged contribution to the ground state. Therefore, they can be neglected. Minimizing
the sum of the Hamiltonians Eq. (\ref{H1}) and Eq. (\ref{Hs}) we find two distinct
states separated by $D(T)=0$. For $D(T)>0$ (i.e. $T>T_{AC}$, ),
the ground state is given by $|\vec{\nabla}_{\perp} u|=|\vec{\delta n}_{\perp}|=0$, which implies that both $\hat{N}$ and $\hat{n}$
point along $\hat{z}$. Thus the system is in the $A$ phase. For $D(T)<0$ (i.e. $T<T_{AC}$),
the ground state shifts to $\partial_s u=\pm\sqrt{-2D(T)/(w-g^2/B)}, \delta n_s=-R\partial_s u, \partial_h u=\delta
n_h=0$. This implies that both $\hat{N}$ and $\hat{n}$ tilt away from $\hat{z}$. However,
the tilting is constrained since $\hat{N}$ and $\hat{n}$ are restricted within $zs$ plane
with $\hat{z}$ in  between the two (this constraint arises from the fact that the last two terms in [\ref{Hs}] will raise the energy of any state in which
$\hat{N}$ and $\hat{n}$ do not lie in this plane above that of an  otherwise identical
state in which they do). In this case the system is
in the $C$ phase. Near the critical point, the magnitudes of the two tilting angles $\theta_N$ and $\theta_n$ are very small
such that approximately, $|\theta_N(T)|=|\partial_s u|$ and $|\theta_n(T)|=|\delta n_s|$.
Therefore, they obey the relation
\begin{eqnarray}
 {\theta_n(T)\over\theta_N(T)}={P\over P+Q}.
 \label{Angle}
\end{eqnarray}

Although Eq. \ref{Angle} is derived through mean field theory, since we are simply minimizing the free energy while ignoring fluctuation effects,
it still holds near the critical point.
This is because the fluctuations of $\vec{\delta n'}_{\perp}$ are always massive. For the same reason we can
set $\vec{\delta n'}_{\perp}=\vec{0}$ in (\ref{Hs}) and use the sum of
(\ref{H1}) and (\ref{Hs}) as our simplified model:
\begin{eqnarray}
H &=& \int d^3r  \left[{ K \over 2}(\nabla^2_{\perp}u)^2 + {B \over
2}(\partial_zu)^2+{D\over 2} (\partial_su)^2
\right.\nonumber\\&-&\left. {g \over 2}(\partial_zu)
(\partial_su)^2+{\gamma\over 2} (\partial_hu)^2+{w
\over 8}
(\partial_su)^4\right.\nonumber\\
&+&\left.\vec{h}\left(\vec{r}\right)\cdot\vec{\nabla}u +
V_p(u-\phi(\vec{r}))\right] \label{CompleteH},
\end{eqnarray}
where the anharmonic terms involving $\partial_h u$ are not included since they are
irrelevant compared to $\gamma(\partial_hu)^2/2$.

To cope with the quenched disorder we employ the replica trick\cite{Geoff}. We assume that the free energy of the system for a specific
realization of the disorder is the same as the one averaged over many realizations. To calculate the averaged
free energy $\overline{F}=\overline{\ln{Z}}$, where $Z$ is the partition function, we use
the mathematical identity $\ln{Z}=\lim_{n\to 0}{Z^n-1\over n}$. When calculating $\overline{Z^n}$, we
can first compute the average over the two random fields $\vec{h}(\vec{r})$ and $\phi(\vec{r})$,
whose statistics have been given earlier. Implementing this procedure gives an replicated Hamiltonian
with the effect of the random fields transformed into  couplings between replicated fields:
\begin{eqnarray}
H[u_{\alpha}] &=& {1 \over 2} \int d^3r \left(\sum^n_{\alpha = 1}
\left[K\left(\partial_s^2u_{\alpha}\right)^2 + B\left(\partial
_zu_{\alpha}\right)^2\right.\right.\nonumber\\
&&\left.\left.- g(\partial
_zu_{\alpha})(\partial_su_{\alpha})^2 + {w \over 4}
(\partial_su_{\alpha})^4 +
D(\partial_su_{\alpha})^2
\right.\right.\nonumber\\
&&+\left.\left.\gamma(\partial_hu_{\alpha})^2\right]-
\sum^n_{\alpha, \beta = 1}\left[{\Delta_t\over T}
\vec{\nabla}_{\perp}u_{\alpha}\cdot\vec{\nabla}_{\perp}u_{\beta}+\right.\right.\nonumber\\
&& \left.\left.{\Delta_c\over T}
(\partial_zu_{\alpha})(\partial_zu_{\beta})+{1\over
T}\Delta_p(u_{\alpha}-u_{\beta})\right]\right). \nonumber\\
\label{Hm}
\end{eqnarray}

We will now analyze this Hamiltonian using the renormalization group.

\section{\label{Sec: RG}Renormalization Group}
Right at the critical point, since the tilt term $(\partial_s u_{\alpha})^2$ vanishes, the Hamiltonian Eq. (\ref{Hm})
becomes very similar to that for a smectic $A$ liquid crystal confined in compressed aerogel; such a system has been described\cite{Karl} as an "$m=1$ Bragg Glass".
For the latter Ref. \cite{Karl} showed that under the renormalization group (RG), the random pinning
$\Delta_p$ is irrelevant, in the RG sense, in spatial dimensions $d<{7\over2}$. Repeating the (virtually identical ) calculation
we find that within the critical region, $\Delta_p$ is also irrelevant in our problem. Furthermore, it is easy to show
that the random compression $\Delta_c$ is also subdominant to $\Delta_t$. Therefore, to construct the RG for the transition
we can use the following truncated Hamiltonian:
\begin{eqnarray}
H[u_{\alpha}] &=& {1 \over 2} \int d^dr \left(\sum^n_{\alpha = 1}
\left[K\left(\partial_s^2u_{\alpha}\right)^2 + B\left(\partial
_zu_{\alpha}\right)^2\right.\right.\nonumber\\
&&\left.\left.- g(\partial
_zu_{\alpha})(\partial_su_{\alpha})^2 + {w \over 4}
(\partial_su_{\alpha})^4 +
D(\partial_su_{\alpha})^2
\right.\right.\nonumber\\
&&+\left.\left.\gamma|\vec{\nabla}_hu_{\alpha}|^2\right]-
\sum^n_{\alpha, \beta = 1}{\Delta_t\over T}
\vec{\nabla}_{\perp}u_{\alpha}\cdot\vec{\nabla}_{\perp}u_{\beta}\right), \nonumber\\
\label{SimpleH}
\end{eqnarray}
where we have adopted the hard analytical continuation with a single
stretch ($z$) axis, a single soft($s$) axis, and $d-2$ hard ($h$) axes.

First we calculate the harmonic propagator. In fourier space the quadratic part of the Hamiltonian can be written as
\begin{eqnarray}
 H={1\over 2}\sum_{q}A_{\alpha\beta}(\vec{q})u_{\alpha}(\vec{q})u_{\beta}(-\vec{q})~,\label{Harmonic}
\end{eqnarray}
where
\begin{eqnarray}
 A_{\alpha\beta}&=&\left(Bq_z^2+\gamma q_h^2+Dq_s^2+Kq_s^4\right)\delta_{\alpha\beta}+{\Delta_t\over T}q_{\perp}^2, \nonumber\\
 \end{eqnarray}
 and
 \begin{eqnarray}
 u_\alpha(\vec{q})&=&{1\over\sqrt{V}}\int d^dr \ \ u_\alpha(\vec{r})e^{-\vec{q}\cdot\vec{r}}.
\end{eqnarray}
The correlation function of this model can be shown to be
\begin{eqnarray}
 \langle u_{\alpha}(\vec{q})u_{\beta}(\vec{q}\,')\rangle=TA_{\alpha\beta}^{-1}(\vec{q})\delta_{\vec{q},-\vec{q}'}\label{prop1}
\end{eqnarray}
where $A_{\alpha\beta}^{-1}$ is the matrix inverse of $A_{\alpha\beta}$. For any $n\times n$ matrix $M_{\alpha\beta}$ of the form
\begin{eqnarray}
M_{\alpha\beta}=a\delta_{\alpha\beta}+b,
\label{Mstand}
\end{eqnarray}
it is straightforward to verify by explicit matrix multiplication that
\begin{eqnarray}
M_{\alpha\beta}^{-1}&=&{1\over a}\delta_{\alpha\beta}+{b\over a(a+bn)}.\label{InverM}
\end{eqnarray}
Since $A_{\alpha\beta}$ is if the form  (\ref{Mstand}), with $a=Bq_z^2+\gamma q_h^2+Dq_s^2+Kq_s^4$ and $b={\Delta_t\over T}q_{\perp}^2$, we obtain, from (\ref{prop1}), upon taking the limit $n\to 0$:
\begin{eqnarray}
 \langle u_{\alpha}(\vec{q})u_{\beta}(-\vec{q})\rangle=TG(\vec{q})\delta_{\alpha\beta}+\Delta_t
q_{\perp}^2G(\vec{q})^2 \label{propagator}
\end{eqnarray}
with
\begin{eqnarray}
G(\vec{q})={1\over Bq_z^2+\gamma q_h^2+Dq_s^2+Kq_s^4}~.
\end{eqnarray}

The momentum shell RG procedure consists of tracing over the short wavelength Fourier modes
of $u_{\alpha}(\vec{r})$ followed by a rescaling of the length.
As usual in the momentum shell RG procedure, we initially restrict wavevectors to lie in a bounded Brillouin zone; here, we will take  our Brillouin zone to be the infinite slab $-\Lambda<q_s<\Lambda$, $-\infty<|\vec{q}_h|<\infty$,
$-\infty<q_z<\infty$, where $\Lambda\sim 1/a$ is an
ultra-violet cutoff, and $a$ is the smectic layer spacing. The displacement field $u_{\alpha}(\vec{r})$
is separated into high and low wave vector parts
$u_{\alpha}(\vec{r})=u_{\alpha}^<(\vec{r})+u_{\alpha}^>(\vec{r})$,
where $u_{\alpha}^<(\vec{r})$ has support in the wave vector space $\Lambda
e^{-d\ell}<q_s<\Lambda$, $-\infty<q_h<\infty$, $-\infty<q_z<\infty$.
We first integrate out $u_{\alpha}^<(\vec{r})$. This integration is done perturbatively in the anharmonic couplings in \ref{SimpleH}; as usual, this perturbation theory can be represented by Feynmann graphs, with the order of perturbation theory reflected by the number of loops in the graphs we consider.  After this perturbative step, we anisotropically rescale lengths,
with $r_s=r'_s e^{\ell}$, $r_h=r'_h e^{\omega_h\ell}$, $r_z=r'_z
e^{\omega_z\ell}$, so as to restore the UV cutoff back to $\Lambda$. This is then followed
by rescaling the long wave length part of the field with $u_{\alpha}^<(\vec{r})=u'_{\alpha}
(\vec{r'})e^{\chi \ell}$. After this procedure we obtain the following RG flow equations to
one-loop order, keeping only lowest order terms in $D$, since we are interested in the critical point where $D$, to leading order in $\epsilon$, vanishes:
\begin{eqnarray}
{d\gamma\over d\ell}&=&\left[(d-4)\omega_h+\omega_z+2\chi+1\right]\gamma~,\\
{dB\over d\ell}&=&\left[(d-2)\omega_h-\omega_z+2\chi+1-{3\over
32\sqrt{2}}g_1\right]B~,\nonumber\\ \label{Bflow}\\
{dK\over d\ell}&=&\left[(d-2)\omega_h+\omega_z+2\chi-3+{1\over
8\sqrt{2}}g_1\right]K,\label{Kflow}\\
{dg\over d\ell}&=&\left[(d-2)\omega_h+3\chi-1+{3\over 16
\sqrt{2}}g_1-{9\over 32 \sqrt{2}}g_2\right]g,\nonumber\\
\label{gflow}\\
{dw\over d\ell}&=&\left[(d-2)\omega_h+\omega_z+4\chi-3-{3\over
8\sqrt{2}}{g_1^2\over g_2}\right]w\nonumber\\&&+\left[{9\over
8\sqrt{2}}g_1-{27\over 32\sqrt{2}}g_2\right]w,\\
{d\Delta_t\over
d\ell}&=&\left[(d-2)\omega_h+\omega_z+2\chi-1+{g_1\over 32\sqrt{2}}\right]\Delta_t,\label{Delta_t}\\
{dD\over d\ell}&=&\left[(d-2)\omega_h+\omega_z+2\chi-1+{9\over
32\sqrt{2}}(g_1-g_2)\right]D\nonumber\\&&+{3\over 8\sqrt{2}}K(g_2-g_1).
\label{GeneralD}
\end{eqnarray}
$g_1$ and $g_2$ are two dimensionless couplings defined by
\begin{eqnarray}
g_1&\equiv& C_{d-1}(g/B)^2\Delta_t
(B\gamma^{2-d}K^{d-7})^{1/2}\Lambda^{2d-7}, ~~~~~~~~\label{g_1def}\\
g_2&\equiv& C_{d-1}(w/B)\Delta_t
(B\gamma^{2-d}K^{d-7})^{1/2}\Lambda^{2d-7}.\label{g_2def}
\end{eqnarray}
where $C_d=S_d/(2\pi)^d$ with $S_d$ the surface area
of a d-dimensional sphere of radius one.
The RG recursion relations (\ref{Bflow}, \ref{Kflow}, \ref{gflow}, \ref{Delta_t}, \ref{GeneralD}) imply that $g_1$ and $g_2$ themselves flow according to the recursion relations
\begin{eqnarray}
 {dg_1\over d\ell}&=&\left[2\epsilon+{21\over 64\sqrt{2}}g_1-{9\over 16\sqrt{2}}g_2\right]g_1,\label{g1flow}\\
 {dg_2\over d\ell}&=&\left[2\epsilon+{63\over 64\sqrt{2}}g_1-{27\over 32\sqrt{2}}g_2-{3\over 8\sqrt{2}}{g_1^2\over g_2}\right]g_2,\nonumber\\
\label{g2flow}
\end{eqnarray}
where $\epsilon\equiv 7/2-d$. These two flow equations are independent of
the rescaling factors $\omega_{h,z}$ and $\chi$ since $g_{1,2}$ are dimensionless.

\begin{figure}
 \includegraphics[width=0.4\textwidth]{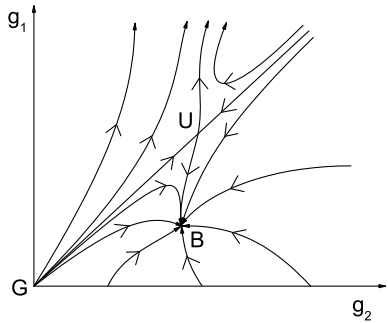}
 \caption{\label{fig: Figure2}Graphical RG flow of the dimensionless couplings $g_1$
and $g_2$, in the $D=0$ plane,  for $d<{7\over2}$. The fixed point $U$ $({128\sqrt{2}\over 15}\epsilon, {128\sqrt{2}\over 15}\epsilon)$
was previously found in \cite{Karl}, and controls $m=1$ BG phase. Both the Gaussian fixed point G $(0, 0)$ and the uniaxial fixed point $U$
are unstable. The biaxial fixed point $B$ $({128\sqrt{2}\over 27}\epsilon, {512\sqrt{2}\over 81}\epsilon)$ is stable, and controls the AC transition in biaxially disordered media.}
\end{figure}

To discuss the critical behavior it is convenient, but not necessary, to make a special choice of $\omega_{z,h}$ such that $\gamma$, $B$, and $K$
are fixed at their bare values. This choice is given by
\begin{eqnarray}
 \omega_h&=&2-{1\over 16\sqrt{2}}g_1,\label{omegah}\\
 \omega_z&=&2-{7\over 64\sqrt{2}}g_1,\label{omegaz}\\
 \chi&=&{5-2d\over 2}+{4d-9\over 128\sqrt{2}}g_1.\label{chi}
\end{eqnarray}
After this choice Eq. (\ref{GeneralD}) becomes
\begin{eqnarray}
 {dD\over d\ell}&=&\left[2+{5g_1\over 32\sqrt{2}}-{9g_2\over
  32\sqrt{2}}\right]D+{3\over 8\sqrt{2}}K(g_2-g_1).\nonumber\\
\label{D}
\end{eqnarray}
For $d<7/2$, there are three fixed points to Eqs. (\ref{g1flow}, \ref{g2flow}, \ref{D}). All three lie on the ``critical surface"\cite{Lubensky}, which separates flows that asymptotically run to large, positive $D$ from those that run to large, negative $D$. More physically, this surface is the phase boundary between the A and the C phase.
The RG flow on this critical surface, projected on to  the $g_{1,2}$ parameter space, is illustrated in Fig. \ref{fig: Figure2}.

This flow is topologically identical to that for the AC transition in a clean system in an external field in spatial dimensions
$d=3-\epsilon$,  for  $\epsilon\ll 1$ \cite{GP}. This is scarcely surprising, since our problem is simply a disordered version of that problem.

The Gaussian fixed point G, given by $g_1^*=0$, $g_2^*=0, D^*=0$ is trivial, and unstable. Hence, it does not control the AC transition.

The uniaxial fixed point U, given by $g_1^*=g_2^*={128\sqrt{2}\over 15}\epsilon+O(\epsilon^2), D^*=0$,  is stable only
within the parameter subspace $g_1=g_2, D=0$. This fixed point was first found in the study of smectics in uniaxially compressed aerogel\cite{Karl}, and controls the phase reference\cite{Karl} calls the ``$m=1$ Bragg Glass". In that system, the constraint $g_1=g_2, D=0$ is enforced by the symmetry that the Hamiltonian is invariant under any rigid
rotation of the liquid crystal about the compressed direction.
This fixed point controls the power law anomalous elasticity of the $m=1$ BG.

In our problem, since the rotation invariance
is destroyed by the stretching of the aerogel, the conditions  $g_1=g_2$ and $D=0$ are not generally obeyed.
Thus, we expect that the critical behavior is controlled by a different fixed point with $g_1^*\neq g_2^*$,
which we call the biaxial fixed point, and  denote as $B$. In fact, the fixed point $B$ must satisfy $g_1^*<g_2^*$, as required by the stability of the model Eq. (\ref{CompleteH}).
To see this we rewrite the model as
\begin{eqnarray}
H &=& \int d^3r  \left[{ K \over 2}(\nabla^2_{\perp}u)^2 + {B \over
2}\left[\partial_zu-\left(g\over 2B\right)\left(\partial_s u\right)^2\right]^2
\right.\nonumber\\&&\left. +{D\over 2} (\partial_su)^2+{\gamma\over 2} (\partial_hu)^2+{1
\over 8}\left(w-{g^2\over B}\right)
(\partial_su)^4\right.\nonumber\\
&&\left.+\vec{h}\left(\vec{r}\right)\cdot\vec{\nabla}u +
V_p(u-\phi(\vec{r}))\right].
\end{eqnarray}
Clearly, for the Hamiltonian to have a well defined minimum at $u=0$ the coefficient of $(\partial_s u)^4$ must be
positive, which implies $g_1<g_2$. This stability condition should be preserved under the RG and, indeed, satisfied
by fixed point B.
To the first order in $\epsilon$, the fixed point B is given by
\begin{eqnarray}
g_1^*&=&{128\sqrt{2}\over 27}\epsilon+O(\epsilon^2), \label{g_1}\\
g_2^*&=&{512\sqrt{2}\over 81}\epsilon+O(\epsilon^2), \label{g_2}\\
D^*&=&-{8K\over 27}\epsilon+O(\epsilon^2)~~~~~~. \label{DFix}
\end{eqnarray}
It is easy to show that this fixed point  is stable in two directions (to leading order in $\epsilon$, the  $g_{1,2}$ plane),  and unstable in a single direction (to leading order in $\epsilon$, the
$D$ direction). This stability/instability structure is precisely that associated in the standard theory of critical phenomenon\cite{Ma's book} with a fixed point controlling a phase transition. Therefore, the fixed point B controls the critical behavior of the AC transition.

\section{\label{Sec: Anomalous Elasticity}Anomalous Elasticity within the Critical Region}

It has long been known  that both pure\cite{GP}  and disordered\cite{RT} smectics exhibit anomalous elasticity, which is a novel phenomenon found, thus far, only in liquid crystals, fluctuating membranes, and at phase transitions\cite{Ma's book, eta}.
In smectics, the anomalous elasticity is characterized by wavevector dependent compression modulus $B$ and bend
modulus $K$ at large length scales. The origin of this exotic behavior is the non-linear coupling of  fluctuations. For a smectic in an isotropic environment (either clean or disordered), rotation invariance symmetry, which
is  spontaneously broken, prohibits certain
terms (e.g., $|\vec{\nabla}_{\perp}u|^2$) in the elastic model, and, hence, the system is soft in the $\perp$ directions. This softness leads to big fluctuations.
As a result, the anharmonic terms in the model become nontrivial and the conventional harmonic theory breaks down. In the presence of
quenched disorder, since the fluctuations are bigger, the anharmonic effect becomes even more important. Thus, in contrast to the marginally weak anomalous elasticity
in pure smectics with logarithmic wavevector dependence, disordered smectics show a strong one with power law
wavevector dependence.

In our problem, since rotation invariance is broken, in general all $\perp$ directions are hard. However,
near the critical point (i.e., at length scales shorter than the correlation lengths), when $\left(\partial_s u_{\alpha}\right)^2$
is subdominant to $\left(\partial_s^2 u_{\alpha}\right)^2$ and hence negligible, the softness in the $s$ direction is recovered.
Indeed, in this case the elastic theory of our system becomes very similar to that for the $m=1$ BG phase of a smectic in uniaxially compressed aerogel\cite{Karl}.
Therefore, we expect similar anomalous behavior. However, the anomalous exponents
will be different, since the system  flows under the RG to a different stable fixed point,
as illustrated in Fig. \ref{fig: Figure2}.

To get an estimate of how important the anharmonic effect is, we calculate the
graphical corrections to the compression modulus $B$ using standard perturbation theory.
The one-loop graphical corrections to $B$ are given by
\begin{eqnarray}
 \delta B&=&-{g^2\over 2}\int d^dp \left[T G^2(\vec{p})+2\Delta_tp_s^2G^3(\vec{p})\right]p_s^4\nonumber\\
         &\approx& {C_{d-1}\Theta_{d-1}\over (2d-7)}g^2\Delta_t\left(\gamma^{2-d}K^{d-7}\over B\right)^{1\over 2}L^{7-2d},~~~
\label{pert}
\end{eqnarray}
where $L^{-1}$ is the infrared cutoff of $q_s$; $L$ will typically be of order the system's length in the soft direction, for systems that are much shorter in the soft than the hard or $z$ directions.  For $d<7/2$, $\delta B$ grows with $L$ and becomes comparable to $B$ at $L=\xi_s^N$,
the ``nonlinear crossover length".  This length $\xi_s^N$ can be calculated by equating $|\delta B|=B$, which gives
\begin{eqnarray}
 \xi_s^N=\left[(7-2d)B^{3\over 2}\gamma^{d-2\over 2}K^{7-d\over 2}\over C_{d-1}\Theta_{d-1}g^2\Delta_t\right]^{1\over 7-2d},
\end{eqnarray}
where $\Theta\equiv\Gamma(d/2)\Gamma(3-d/2)/2$. The corresponding lengths along the $h$ and $z$ axes can be obtained for looking for the wavenumbers $q_h^N\equiv1/\xi_h^N$ and $q_z^N\equiv1/\xi_z^N$ at which the $B$ and $\gamma$
terms in the propagator $G(\vec{p})$ become comparable to the $K$
term evaluated at $ \xi_s^N$. This gives:
$\xi_h^N=\left(\xi_s^N\right)^2/\lambda_h$
and $\xi_z^N=\left(\xi_s^N\right)^2/\lambda_z$, where $\lambda_z\equiv\sqrt{K/B}$ and $\lambda_h\equiv\sqrt{K/\gamma}$.
The conventional harmonic theory only applies when at least one of the length scales being probed in the soft, hard, or $z$ directions is smaller than the corresponding non-linear crossover lengths $\xi_{s,z,h}^N$. To go beyond
these crossover lengths, we employ the RG which has been derived in section \ref{Sec: RG}.

To do this we use the trajectory integral matching formalism\cite{Nelson}, as used in\cite{RT} and many other papers. In this approach, the RG is used
established a connection between a correlation function at small wave vector in the unrenormalized system, and the same correlation function at a larger wavevector in the renormalized system.
While the former cannot be evaluated using the harmonic approximation
theory, the latter can, if we rescale to the point where the rescaled wavevector lies within the harmonic regime just described. Here, we chose the connected disorder averaged correlation function is $C(\vec{q})=\overline{\langle |u(\vec{q})|^2\rangle}-\overline{\langle u(-\vec{q})\rangle\langle u(\vec{q})\rangle}$.

For this correlation function, the connection just described is given by:
\begin{eqnarray}
 &&C(\vec{q}, B, K, \gamma, \Delta_t, g, w)\nonumber\\
 &=&e^{\left[2\chi+(d-2)\omega_h+\omega_z+1\right]\ell}C\left(q_se^\ell, q_ze^{\omega_z\ell}, q_he^{\omega_h\ell},\right. \nonumber\\
 &&\left.B(\ell), K(\ell), \gamma(\ell), \Delta_t(\ell), g(\ell), w(\ell)\right),
 \label{ConnectionCF}
\end{eqnarray}
where on the right-hand side the prefactor comes from the dimensional and field rescalings.

First we consider special $\vec{q}\,$'s with $q_z=0, q_h=0$. In this case we will choose the rescaling parameter is taken to be $\ell=\ell^*\equiv\ln\left(\Lambda/q_s\right)$,
so that the rescaled $\vec{q}\,$'s lies on  the boundary of the Brillouin zone. Thus, we can accurately evaluate the correlation function on the right-hand side
of Eq. (\ref{ConnectionCF}) using the harmonic approximation, and obtain
\begin{eqnarray}
 &&C(\vec{q}, B, K, \gamma, \Delta_t, g, w)\nonumber\\
 &=&e^{\left[2\chi+(d-2)\omega_h+\omega_z+1\right]\ell^*}{1\over K(\ell^*)\Lambda^4}\nonumber\\
 &=&{1\over K(q_s, q_{z, h}=0)q_s^4}
 \label{L1}
\end{eqnarray}
where
\begin{eqnarray}
 K(q_s, q_{z, h}=0)=K\left(\xi_s^Nq_s\right)^{-\eta_K}
\end{eqnarray}
with
\begin{eqnarray}
 \eta_K={g_1^*\over 8\sqrt{2}}={16\over 27}\epsilon+O(\epsilon^2).
\end{eqnarray}

In the second equality of Eq. (\ref{L1}) we have eliminated $\ell^*$ in favor of $q_s$ and used
\begin{eqnarray}
 K(\ell^*)&=&K\left(\xi_s^N\Lambda\right)^{(d-2)\omega_h+\omega_z+2\chi-3}\times\nonumber\\
          && \left(\xi_s^Nq_s\right)^{(2-d)\omega_h-\omega_z-2\chi+3-{g_1^*\over 8\sqrt{2}}}\nonumber\\
          &=&\left(\xi_s^Nq_s\right)^{-{g_1^*\over 8\sqrt{2}}}\left(\Lambda\over q_s\right)^{(d-2)\omega_h+\omega_z+2\chi-3}
\end{eqnarray}
which is obtained by integrating the flow Eq. (\ref{Kflow}).

The above calculations can be generalized for arbitrary direction of $\vec{q}$ by using a more sophisticated
choice of $\ell^*$, in which we rescale until the propagator $G(\vec{p})$ takes on the same value as its smallest on the Brillouin zone boundary. This leads to the condition:
\begin{eqnarray}
 K(\ell^*)\Lambda^4&=&K(\ell^*)(q_se^{\ell^*})^4+B(\ell^*)(q_ze^{\omega_z\ell^*})^2+\nonumber\\
                   &&\gamma(\ell^*)(q_he^{\omega_h\ell^*})^2.
                   \label{ell*}
\end{eqnarray}
This choice again ensures that the correlation function on the right-hand side of Eq. (\ref{ConnectionCF}) can be accurately
evaluated using the harmonic approximation. Clearly, the solution of this equation reduces to $\ell^*=\ln\left(\Lambda/q_s\right)$
for the special case $q_h=0, q_z=0$. In general, the solution is
\begin{eqnarray}
 e^{\ell^*}=\left(\Lambda\over q_s\right)f\left(X, Y\right)
 \label{GeneralEll}
\end{eqnarray}
where we have defined the scaling variables
\begin{eqnarray}
X\equiv {q_z\xi_z^N\over \left(q_s\xi_s^N\right)^{\zeta_z}}~~,
\label{Xdef}
\end{eqnarray}
and
\begin{eqnarray}
Y\equiv {q_h\xi_h^N\over \left(q_s\xi_s^N\right)^{\zeta_h}}~~.
\label{Ydef}
\end{eqnarray}
The universal anisotropy exponents
 $\zeta_z$ and $\zeta_h$ are given by
\begin{eqnarray}
\zeta_z&=&2-{(\eta_B+\eta_K)\over 2},\\
\zeta_h&=&2-{\eta_K\over 2}.
\end{eqnarray}
In equation (\ref{GeneralEll}), $f(X, Y)$
is a scaling function. Its asymptotic behaviors can be deduced by noting that, for $X\ll1$, $Y\ll1$, it should recover the result for $q_z=q_h=0$, while for $X\gg 1, X^{\zeta_h}\gg Y^{\zeta_z}$, $\ell^*$ should only depend on $q_z$, and for  $Y\gg1, Y^{\zeta_z}\gg X^{\zeta_h}$, $\ell^*$ should only depend on $q_h$. These conditions  force $f(X,Y)$ to have the asymptotic forms:
\begin{eqnarray}
 f(X, Y)=\left\{\begin{array}{ll}
 1, ~~&X\ll 1,~~ Y\ll 1,\\
 X^{-{1\over\zeta_z}}, ~~&X\gg 1, ~~X^{\zeta_h}\gg Y^{\zeta_z},\\
 Y^{-{1\over\zeta_h}}, ~~&Y\gg 1, ~~Y^{\zeta_z}\gg X^{\zeta_h}.
 \end{array}
 \right.
\end{eqnarray}
Using $\ell^*$ as given by Eq. (\ref{GeneralEll}) and repeating our earlier calculation for the $q_h=q_z=0$ case, we obtain the
anomalous elasticity for $B$, $K$ and $\Delta_t$ for an arbitrary direction of $\vec{q}$:
\begin{eqnarray}
B(\vec{q})&=&B\left(\xi_s^Nq_s\right)^{\eta_B}(f\left(X, Y\right))^{-\eta_B} \nonumber \\ &\equiv&
\left(\xi_s^Nq_s\right)^{\eta_B}f_B\left(X, Y\right)
\label{AnomalousB}
\end{eqnarray}
\begin{eqnarray}
 K(\vec{q})&=&K\left(\xi_s^Nq_s\right)^{-\eta_K}(f\left(X, Y\right))^{\eta_K} \nonumber \\
 &\equiv& \left(\xi_s^Nq_s\right)^{-\eta_K}f_K\left(X, Y\right)
 \label{AnomalousK}
 \end{eqnarray}
 \begin{eqnarray}
 \Delta_t(\vec{q})&=&\Delta_t\left(\xi_s^Nq_s\right)^{-\eta_t}(f\left(X, Y\right))^{\eta_t} \nonumber \\
&\equiv& \left(\xi_s^Nq_s\right)^{-\eta_t}f_t\left(X, Y\right)
 \label{AnomalousDelta}
\end{eqnarray}
where we have defined the $B$, $K$, and $\Delta_t$ scaling functions:
\begin{eqnarray}
f_{B}(X,Y)\equiv B (f\left(X, Y\right))^{-\eta_B}~~,
\end{eqnarray}
\begin{eqnarray}
f_{K, t}(X, Y)\equiv (K, \,\Delta_t)(f\left(X, Y\right))^{\eta_{K,t}}~~.
\end{eqnarray}
Here, the exponents $\eta_B$ and $\eta_t$ are given by:
\begin{eqnarray}
 \eta_B={3\over 32\sqrt{2}}g_1^*={4\over 9}\epsilon+O(\epsilon^2),\\
 \eta_t={1\over 32\sqrt{2}}g_1^*={4\over 27}\epsilon+O(\epsilon^2).
\end{eqnarray}
In deriving Eq. (\ref{AnomalousDelta}), we have used the correlation function $\overline{\langle
u(-\vec{q})\rangle\langle u(\vec{q})\rangle}=\Delta_t q_s^4G^2(\vec{q})$ instead of $C(\vec{q})$ ,
since the latter has no dependence on $\Delta_t$.

In the physical dimension $d=3$, to  leading order in $\epsilon$, $\eta_B=0.2222$, $\eta_K=0.2963$, and $\eta_t=0.07407$ . These results are quantitatively reliable, since $\epsilon={1\over 2}$, which is small. Since both $\eta_B$ and $\eta_k$ are positive, the compression modulus $B$ and the bend modulus $K$ vanish and diverge, respectively,  as $\vec{q}\to \vec{0}$. Physically, this can be understood as follows:

In the presence of layer fluctuations, a compression can be relieved by smoothing out these fluctuations, thus leading to a  vanishing effective compression modulus. Similarly, bending a fluctuating smectic  leads to compression of the layers, thus leading to a diverging effective bend modulus.

\section{\label{Sec: Critical Behavior}Critical Behavior}
In this section we calculate the critical exponents for the transition following standard RG procedures \cite{Lubensky}. Linearizing the flow Eqs. (\ref{g1flow}, \ref{g2flow}, \ref{D}) around
the fixed point $g_1^*={128\sqrt{2}\over 27}\epsilon+O(\epsilon^2),~
g_2^*={512\sqrt{2}\over 81}\epsilon+O(\epsilon^2),~
D^*={8\over 27}\epsilon+O(\epsilon^2)$, we obtain
\begin{eqnarray}
 {d\delta g_1\over d\ell}&=&{14\epsilon\over 9}\delta g_1-{8\epsilon\over 3}\delta g_2,\\
 {d\delta g_2\over d\ell}&=&{8\epsilon\over 3}\delta g_1-4\epsilon\delta g_2,\\
 {d\delta D\over d\ell}&=&\lambda_D\delta D+{3\over 8\sqrt{2}}\delta g_1-{3\over 8\sqrt{2}}\delta g_2,
\end{eqnarray}
where
\begin{eqnarray}
 \lambda_D=2-{28\over 27}\epsilon+O(\epsilon^2).
\end{eqnarray}
These recursion relations have the solution:
\begin{eqnarray}
 \delta g_1&=&C_1e^{-2\epsilon\ell}+C_2e^{-{4\over9}\epsilon\ell},\label{deltag1}\\
 \delta g_2&=&{4\over 3}C_1e^{-2\epsilon\ell}+{3\over 4}C_2e^{-{4\over9}\epsilon\ell},\label{deltag2}\\
 \delta D&=&C_3e^{\lambda_D\ell}+{KC_1\over 16\sqrt{2}}e^{-2\epsilon\ell}-{3KC_2\over 64\sqrt{2}}e^{-{4\over 9}\epsilon\ell},
 \label{deltaD}
\end{eqnarray}
where the $C$'s are determined by the bare values of the coefficients in the Hamiltonian Eq. (\ref{SimpleH}).

Note that  $\delta D$ flows to $\infty$ for positive $C_3>0$, to $-\infty$ for $C_3<0$,
and to zero for $C_3=0$. Thus positive $C_3$ corresponds to the A phase, negative $C_3$ to the C phase, and $C_3=0$ to the AC transition.
In terms of the bare values of $g_{1,2}$ and $D$, the AC transition occurs when $C_3=0$; a little algebra shows that this happens when
\begin{eqnarray}
 D(T_{AC})={3K\over16\sqrt{2}}\left(g_2-g_1\right),
\end{eqnarray}
which describes a plane in the $g_1$-$g_2$-$D$ parameter space. Unlike in  mean field
theory, the critical value of $D$, and hence the critical temperature $T_{AC}$ depend on both $g_1$ and $g_2$.

$C_3$ measures how far the system is away from the critical point. Assuming that $C_3$ is a smooth, analytic function of temperature near $T_{AC}$ (and we have no reason to assume otherwise), then near the critical point, $C_3\propto t\equiv {T-T_{AC}\over T_{AC}}$. This is the sole input we will need in our analysis to get the critical dependence of all quantities on $t$.

For $t\to 0^+$, the correlation lengths of the original system and those of the rescaled system are connected by
\begin{eqnarray}
 \xi_s&=&e^{\ell}\xi_s(\ell),\\
 \xi_h&=&e^{\omega_h\ell}\xi_h(\ell),\\
 \xi_z&=&e^{\omega_z\ell}\xi_z(\ell).
\end{eqnarray}
This allows us to calculate the singular $t$-dependence of the correlation lengths.
Under the RG, according to flow Eqs. (\ref{deltag1}, \ref{deltag2}, \ref{deltaD}) $\delta g_1\to 0$, $\delta g_2\to 0$,
and $\delta D$ grows as $\delta D(\ell)\propto te^{\lambda_D\ell}$ . At
\begin{eqnarray}
\ell=\ell^*\equiv -{1\over\lambda_D}\ln{t}~~,
\label{ell^*}
\end{eqnarray}
 $\delta D(\ell)$ is of order one, and the rescaled system becomes non-critical with finite
$\xi_{s,h,z}(\ell)$. Plugging $\ell=\ell^*$ into the above equations we obtain
\begin{eqnarray}
 \xi_s\propto t^{-\nu_s},\\
 \xi_z\propto t^{-\nu_z},\\
 \xi_h\propto t^{-\nu_h},
\end{eqnarray}
with
\begin{eqnarray}
 \nu_s&=&{1\over \lambda_D}={1\over 2}+{7\over 27}\epsilon+O(\epsilon^2),\label{nus}\\
 \nu_z&=&\zeta_z\nu_s=1+{7\over 27}\epsilon+O(\epsilon^2),\label{nuz}\\
 \nu_h&=&\zeta_h\nu_s=1+{10\over 27}\epsilon+O(\epsilon^2).\label{nuh}
\end{eqnarray}

The temperature dependence of the order parameter $\theta_N$, the angle between $\hat{N}$ and $\hat{z}$, can be calculated in a similar way.
For $t\to 0^-$, $\theta_N$ is small and approximately $\theta_N=|\partial_s u|$. $|\partial_s u|$ in the original system and that in the
rescaled system are connected by
\begin{eqnarray}
 |\partial_s u|=|\partial_s u(\ell^*)|e^{(-1+\chi)\ell^*}.
 \label{OP}
\end{eqnarray}
$|\partial_s u(\ell^*)|$ can be calculated using mean field theory since the rescaled system is deep into the $C$ phase, and given by
\begin{eqnarray}
 |\partial_s u(\ell^*)|=\sqrt{-2D(\ell^*)\over {w(\ell^*)-{g^2(\ell^*)\over B(\ell^*)}}}.
 \label{RenormalizedOP}
\end{eqnarray}
Since the dimensionless coupling $g^*_{1,2}$ are finite constants near the RG fixed point, and we have chosen our RG rescaling factors so that, $K(\ell)$, $B(\ell)$, and $\gamma(\ell)$ are kept fixed at their bare values,
we have, from the definitions Eqs. (\ref{g_1def}, \ref{g_2def}), $w(\ell^*)\sim {1\over \Delta_t(\ell^*)}$ and $g^2\sim {1\over \Delta_t(\ell^*)}$. Furthermore, $D(\ell^*)$ is of order one.
Therefore, Eq. (\ref{RenormalizedOP}) leads to $|\partial_s u(\ell^*)|\sim \sqrt{\Delta_t(\ell^*)}$. Substituting
this result into Eq. (\ref{OP}), using the recursion relation equation (\ref{Delta_t}) for $\Delta_t$, and using  equation (\ref{ell^*}) for $\ell^*$, we get
\begin{eqnarray}
 |\partial_s u|\sim |t|^{{\nu_s\over 2}\left[(2-d)\zeta_h-\zeta_z-7+4d-\eta_B-\eta_t-(d-3)\eta_K\right]}
\end{eqnarray}
which implies
\begin{eqnarray}
 \beta
      &=&{\nu_s\over 4}\left[4d-10+(5-d)\eta_K-\eta_B-2\eta_t\right]\nonumber\\
      &=&{1\over 2}-{2\over 9}\epsilon+O(\epsilon^2).
\end{eqnarray}
The first and second equalities are respectively the exact scaling relation (\ref{ExactBeta}) and the $\epsilon$-expansion result (\ref{betaeps}) quoted in the introduction.

The specific heat exponent can be obtained by similar techniques\cite{Nelson}. The disorder averaged free energy of
the system is invariant under the RG, that is,
\begin{eqnarray}
 \overline{F}&=&V_r\overline{f}_r\left[K(\ell^*), B(\ell^*), \gamma(\ell^*), D(\ell^*), \Delta_t(\ell^*)\right]\nonumber\\
  &~& + \Delta F(\ell^*),\label{Free E RG}
\end{eqnarray}
where $V_r$ and $\overline{f}_r$ are the volume and disorder averaged free energy density of the rescaled system, respectively, and $ \Delta F(\ell^*)$ represents the contribution to the free energy from the degrees of freedom already integrated out in the RG process. We will focus in what follows on the first term on the right hand side of equation  (\ref{Free E RG}), since it scales the same way as the full free energy.
If renormalize for the same renormalization group time $\ell^*$ that we used above to calculate the correlation lengths, we will again have renormalized out of the critical region, which will permit us to calculate the free energy in mean field theory. With this choice,
$V_r$ is related to the volume of the original system by $V_r=V_0e^{-\left[\left(d-2\right)\omega_h+\omega_z+1\right]\ell^*}$.
Furthermore, because we have renormalized out of the critical regime, $f_r$ only depends on $K(\ell^*)$, $B(\ell^*)$, $\gamma(\ell^*)$, $D(\ell^*)$, and $\Delta_t(\ell^*)$
since it can be evaluated by ignoring the anharmonic terms. The only singular temperature
dependence of $\overline{f}_r(\ell^*)$ comes from $\Delta_t(\ell^*)$, since $K(\ell^*)$, $B(\ell^*)$, and $\gamma(\ell^*)$ are fixed
at their bare values and $D(\ell^*)$ is of order one. This singular temperature dependence can be
calculated by noting that $\overline{f}_r$ is linear in $\Delta_t(\ell^*)$; that is
\begin{eqnarray}
 {\partial \overline{f}_r(\ell^*)\over \partial \Delta_t(\ell^*)}&=&\lim\limits_{n\to 0}{1\over (2\pi)^dn}\int d^dq \ \ q_s^2\sum_{\alpha\beta}\overline{\langle u_{\alpha}(\vec{q})u_{\beta}(-\vec{q})\rangle}\nonumber\\
 &=&\mbox{constant}.
\end{eqnarray}
Therefore,
\begin{eqnarray}
 \overline{F}&\propto& V_0e^{-\left[\left(d-2\right)\omega_h+\omega_z+1\right]\ell^*}\left[\Delta_t(\ell^*)+\mbox{cons.}\right]\nonumber\\
  &\propto& |t|^{\nu_s\left[(d-2)\zeta_h+\zeta_z-1+\eta_K-\eta_t\right]}
  \label{Free E crit}
\end{eqnarray}
where in the second line we have used the solution of the recursion relation (\ref{Delta_t}) and the expression (\ref{ell^*}) relating $\ell^*$ to the reduced temperature $t$.

Because the specific heat is the second temperature derivative of the free energy, the power of $t$ on the right hand side of the bottom line of equation (\ref{Free E crit}) is just $2-\alpha$, where $\alpha$ is the usual specific heat critical exponent. Thus,
\begin{eqnarray}
 \alpha
       &=&2-{\nu_s\over 2}\left[4d-6+(3-d)\eta_K-2\eta_t-\eta_B\right]\nonumber\\
       &=&{2\over 9}\epsilon+O(\epsilon^2).
\end{eqnarray}
The first and second equalities are respectively the exact scaling relation (\ref{ExactAlpha}) and the $\epsilon$-expansion result (\ref{alphaeps}) quoted in the introduction.

Once we probe the system at length scales longer than the correlation lengths $\xi_{s,h,z}$ in the corresponding directions, the elastic constants and disorder variances become wavevector independent. This can be seen either by repeating the RG analysis just done above for small $D$ for large $D$; alternatively, one could repeat the perturbation calculation equation (\ref{pert}) with a nonzero $D$, and find that all perturbative corrections now converge in the infra-red.

Instead of wavevector dependent, however, all of the elastic constants and disorder variances
now become
 the temperature dependent. We can obtain this temperature dependence by matching.

 For example, the temperature dependence of $B$ can be obtained by matching, right at $q_s=\xi_s^{-1}$, the wavevector dependence of $B(q_s, q_z=0, \vec{q}_h=\vec{0})$ for $q_s\gg\xi_s^{-1} $ given   in equation (\ref{AnomalousB}) onto the wavevector independent behavior we expect for
$q_s \ll \xi_s^{-1}$. Doing so, we obtain $B(T)=B(\xi_s^N/\xi_s(T))^{\eta_B}$.


We can obtain $D(T)$ using
the definition of the correlation length $\xi_s$, which implies $K(\vec{q})q_s^2=D(T)$ at $q_s=\xi_s^{-1}, q_z=0, q_h=0$; this implies  $D(T)\propto(\xi_s^N/\xi_s)^{-\eta_K}\xi_s^{-2}$.

Similar arguments give $K(T)$, and the disorder variances $\Delta_{t,c}(T)$, where the needed {\it critical} renormalization of the compression disorder variance $\Delta_c(\vec{q})$ is computed in appendix (\ref{sec: AnomalousElasticityIrrelevant}). Summarizing all of these results:
\begin{eqnarray}
B(T) &=& B(\xi_s^N/\xi_s)^{\eta_B}\propto
(T-T_{AC})^{\eta_B\nu_s}~,
\label{critB}
\end{eqnarray}
\begin{eqnarray}
K(T) &=& K(\xi_s^N/\xi_s)^{-\eta_K}\propto
(T-T_{AC})^{-\eta_K\nu_s},
\nonumber \\
\label{critK}
\end{eqnarray}
\begin{eqnarray}
D(T) &=& D(\xi_s^N/\xi_s)^{2-\eta_k-{1\over\nu_s}}\propto
(T-T_{AC})^{(2-\eta_K)\nu_s}~,\nonumber\\
\label{critD}
\end{eqnarray}
\begin{eqnarray}
\Delta_{t}(T) &=&
\Delta_{t}(\xi_s^N/\xi_s)^{-\eta_{t}}\propto
(T-T_{AC})^{-\eta_{t}\nu_s}~,\nonumber\\
\label{critDt}
\end{eqnarray}
\begin{eqnarray}
 \Delta_{c}(T) &=&
 \Delta_{c}(\xi_s^N/\xi_s)^{-\eta_{c}}\propto
(T-T_{AC})^{-\eta_{c}\nu_s}~.\nonumber\\
\label{critDc}
\end{eqnarray}

These
temperature dependences are needed to predict the singular dependence of the light scattering in the A and C phases near the critical point, as will be shown in section (\ref{Sec: scat}).

\section{\label{Sec: phases}Properties of the A and C Phases}

The analysis of the preceding two sections has focused on the critical region near the AC transition. What about the properties of the A and C phases themselves in this random biaxial environment?

These prove to be quite interesting and novel as well: we find that {\it both} the A and C phases behave, at long distances, like an XY ferromagnet in a random field\cite{Fisher}. This should be contrasted to the behavior of this system in {\it uniaxial} disordered environments, in which case the A phase also looks like an XY model in a random field, but the C phase belongs to the far more disordered ``m=1 Bragg Glass" universality class\cite{Karl}.

We begin with the A phase. At long length scales $r_s\ll\xi_s$, $r_z\ll\xi_z$, and $r_h\ll\xi_h$)(i.e., $D>0$), the effective Hamiltonian for this phase is:
\begin{eqnarray}
H &=& \int d^dr  \left[{B(T) \over
2}(\partial_zu)^2+{D(T)\over 2} (\partial_su)^2+{\gamma\over 2} (\partial_hu)^2
\right.\nonumber\\
&&\left.+V_p(u-\phi(\vec{r}))\right] \label{RFXY}.
\end{eqnarray}
Note that we have dropped the bend term $(\partial_s^2 u)^2$, because it is less important than the $(\partial_s u)^2$; near the transition, we could not do this, since $D(T)$ vanished there. But in the A phase, since $D>0$, we have this term, and it dominates the bend term at long length scales as defined above.

An additional effect of this extra term is that it suppresses fluctuations of $u$ sufficiently that the $g$ and $w$ anharmonicities in (\ref{H1}) are also
irrelevant, as are the random tilt terms.
Hence, the only relevant anharmonicity is that implicitly contained in the random pinning potential $V_p(u-\phi(\vec{r}))$, which we remind the reader is a periodic function with period $a$.

Note that neither the compression modulus $B(T)$ nor the coefficient $D(T)$ in the Hamiltonian Eq. (\ref{RFXY}) are given by their bare values. Instead,
both are renormalized by the critical fluctuations and become non-trivially temperature dependent. They are given by Eqs. (\ref{critB}, \ref{critD}).



Like the model for the smectic A phase in {\it uniaxially}\cite{Karl} stretched aerogel, this model
is effectively a random field XY model ($RFXY$),  with an anisotropic spin wave stiffness.

To show this, we rescale  lengths as follows:
\begin{eqnarray}
r_s=r_s'\sqrt{D/B}~~, ~r_h=r_h'\sqrt{\gamma/B}~~
\label{rescale r},
\end{eqnarray}
and replace $u$ by a rescaled variable
\begin{eqnarray}
\theta(\vec{r}\,')\equiv q_0u(\vec{r})~~
\label{rescale u},
\end{eqnarray}
where $q_0\equiv2\pi/a$ is the reciprocal lattice constant. This latter transformation has the property that the physical
state of the system is now invariant under the transformation:
 \begin{eqnarray}
 \theta\rightarrow\theta+2\pi n~~,
 \label{theta}
\end{eqnarray}
where $n$ is any integer, since the smectic state is invariant under translations by any integral number $n$ of layer spacings $a$. This is precisely the symmetry obeyed by the angle variable $\theta_{XY}$ giving the angle between the local spins and some reference direction in a ferromagnetic XY model. And, indeed, in  the rescaled variables, the model  becomes:

\begin{eqnarray}
H =\int d^dr^\prime  \left[{K_{XY} \over
2}|\vec{\nabla} '\theta|^2+V^{XY}_p(\theta-\phi_{XY}(\vec{r}))\right],~~~~~
\label{RFXY2}
\end{eqnarray}
where$V^{XY}_p(\theta) \equiv V_p({a\theta\over2\pi})$ is a periodic function with period $2\pi$,  $\phi_{XY}(\vec{r}) \equiv {2\pi \phi(\vec{r})\over a}$ is a short-range correlated random variable uniformly distributed between $0$ and $2\pi$, and, in spatial dimension $d=3$,  the spin wave stiffness $K_{XY}={a^2\sqrt{D\gamma}\over4\pi^2}$.

This Hamiltonian (\ref{RFXY2}) is easily recognized\cite{easyRFXY} as just the long-wavelength Hamiltonian for a ferromagnetic XY model in a random field,
which has been extensively studied\cite{Fisher}.
Fisher\cite{Fisher} showed that , in this model, the spatio-temporally Fourier transformed two point correlations of $\theta$ are given by $\overline{\langle \theta(\vec{q}\,')\theta(-\vec{q}\,')\rangle}={C(d)\over q'^d}$, where $C(d)$ is an constant of order one and $C(3)\approx 1.10\pi^2$.   Using the rescaling (\ref{rescale u}) with this result then implies:
$\overline{\langle u(\vec{q}\,')u(-\vec{q}\,')\rangle}={C(d)\over q'^dq_0^2}$.  Then using $u(\vec{q}\,')=\sqrt{V'/V}u(\vec{q})$, where $V$ and $V'$ are respectively the volumes of the system in the original and primed coordinate systems, and rescaling $\vec{q}\,'$ back to $\vec{q}$ using the inverse of the rescalings (\ref{rescale r}), we find, in $d=3$,
\begin{eqnarray}
 \overline{\langle u(\vec{q})u(-\vec{q})\rangle}={1.10\pi^2\sqrt{BD\gamma}\over\left(Bq_z^2+Dq_s^2+\gamma q_h^2\right)^{3\over 2}q_0^2}~~
 \label{uqRFXY}.
\end{eqnarray}

The real space correlation function
\begin{eqnarray}
C(\vec{r})\equiv\overline{\langle [u(\vec{r})-u(\vec{0})]^2\rangle}
\end{eqnarray}
can be calculated using a similar strategy.
The real space correlation function governed by Hamiltonian (\ref{RFXY2}) has also been derived by Fisher\cite{Fisher}, and is, in $d=3$, $\overline{\langle [\theta(\vec{r}\,')-\theta(\vec{0})]^2\rangle}=1.10\ln\left(r'/b'\right)$,
where $b'$ is the cutoff length in $\vec{r}\,'$ space. Then rescaling  $\theta$ and $\vec{r}\,'$ back to $u$ and $\vec{r}$ , respectively, using the rescalings described above, we obtain
\begin{eqnarray}
 C(\vec{r})\thickapprox {1.10\over q_0^2}\ln{\sqrt{\left(r_s\over b_s\right)^2+\left(r_h\over b_h\right)^2+\left(r_z\over b_z\right)^2}}~~~\label{C_A}
\end{eqnarray}
where $b_{s,h,z}$ are respectively cutoffs along $\hat{s}$, $\hat{h}$, and $\hat{z}$ in $\vec{r}$ space, and related to $b'$ by the
rescalings (\ref{rescale r}), which give $b_s=b'\sqrt{D/B}$, $b_h=b'\sqrt{\gamma/B}$, and $b_z=b'$.

The physically more important correlation is $F_n(\vec{r})=\overline{\langle e^{inq_0[u(\vec{r})-u(0)]}\rangle}$, which can be experimentally measured by  X-ray scattering, as will be  discussed in next section. In the approximation that the fluctuations of $u$ are Gaussian, which Fisher\cite{Fisher} has shown to give the correct power law decay of the XY analog of  $F_n(\vec{r})$ to leading order in $\epsilon\equiv 4-d$,
\begin{eqnarray}
F_n(\vec{r})\thickapprox \exp\left[-{n^2q_0^2\over 2}C(\vec{r})\right].\label{Fn(r)}
\end{eqnarray}
Plugging Eq. (\ref{C_A}) into Eq. (\ref{Fn(r)}) we get
\begin{eqnarray}
 F_n(\vec{r})\thickapprox\left(\sqrt{\left(r_s\over b_s\right)^2+\left(r_h\over b_h\right)^2+\left(r_z\over b_z\right)^2}\right)^{-0.55n^2}
 \label{Fn(r)2}
\end{eqnarray}
i.e., quasi-long-ranged translational order. However, unlike in bulk smectics\cite{Caille}, the power law exponent has no dependence on either the temperature or the quantitative elastic properties
(e.g., the bare values of the elastic coefficients), and is therefore universal. Furthermore, $F_n(\vec{r})$ is isotropic in scaling, unlike the clean case\cite{Caille}.

Now we discuss $\overline{\langle u(\vec{q})u(-\vec{q})\rangle}$ at small $\vec{q}\,$'s in the $C$ phase. Since $D<0$, $u=0$ is not the disorder averaged ground
state. Therefore, we need to first expand the Hamiltonian around the state $\partial_zu=-{Dg\over Bw-g^2}, \partial_hu=0, \partial_s u=\sqrt{-{DB\over Bw-g^2}}$, which is one of the disordered averaged ground states in the $C$ phase, as obtained by minimizing the Hamiltonian Eq. (\ref{CompleteH}) with
the disordered terms excluded. We will denote the displacement of the layer from the ground state position  as $u'(\vec{r})$. Expanding in small $u\,'$,  and ignoring all irrelevant terms, we obtain a Hamiltonian which is essentially the same as Eq. (\ref{RFXY}) but with $u(\vec{r})$ replaced by $u'(\vec{r})$. Thus, we conclude that the universality class of the $C$ phase is also that of the random field XY model, and $\overline{\langle u'(\vec{q})u'(-\vec{q})\rangle}$ is also given by equation (\ref{uqRFXY}). Likewise,  real space correlations are also given by equation (\ref{Fn(r)2}).

\section{\label{Sec: scat}Predictions for Scattering Experiments}

In this section we work out the implications of our theory for both visible light and X-ray scattering experiments.

\subsection{\label{Sec: X-ray}X-ray Scattering}
X-ray scattering experiments provide a direct experimental measure of the density-density correlation function $\overline{\langle\rho(\vec{r})\rho(\vec{r}\,')\rangle}$, where $\rho(\vec{r})$ is the
molecular number  density. This correlation function  is an important quantity
in smectics,  since it reflects the translational order of the system. The X-ray
scattering intensity is related to its Fourier transform  by
\begin{eqnarray}
 I(\vec{q})\propto\int d^dr' d^dr''\, \overline{\langle\rho(\vec{r}\, ')\rho(\vec{r}\, '')\rangle} e^{-i\vec{q}\cdot(\vec{r}\,'-\vec{r}\, '')}.
 \label{Xray1}
\end{eqnarray}
In a smectic, $\rho(\vec{r})$ can be expanded in a Fourier series with period $a$, the layer spacing
between nearest layers, via
\begin{eqnarray}
 \rho(\vec{r})={1\over \sqrt{V}}\sum_{n}\rho_n e^{inq_0\left[z+u(\vec{r})\right]},
 \label{rho}
\end{eqnarray}
where $\rho_n$ is the complex amplitude of the $n_{th}$ harmonic of the density wave. In writing this equation (\ref{rho}), we have included the effect of fluctuations $u(\vec{r})$ in the positions of the layers, but not those of the magnitudes. The latter fluctuations can be shown\cite{Toner's thesis} to be irrelevant, in the RG sense, close to the transition.

Inserting Eq. (\ref{rho}) into Eq. (\ref{Xray1}) we obtain
\begin{eqnarray}
 I(\vec{q})\propto \sum_n |\rho_n|^2\int d^dr \, e^{i(nq_0\hat{z}-\vec{q})\cdot\vec{r}}F_n(\vec{r}),
 \label{Xray2}
\end{eqnarray}
where
\begin{eqnarray}
 F_n(\vec{r})=\overline{\langle e^{inq_0[u(\vec{r})-u(\vec{0})]}\rangle}.
\end{eqnarray}
In deriving Eq. (\ref{Xray2}), we have changed variables of integration from $\vec{r}\,'$
to $\vec{r}=\vec{r}\,'-\vec{r}\,''$ and used the fact that, for homogeneous systems,
\begin{eqnarray}
\overline{\langle e^{inq_0[u(\vec{r}\,')-u(\vec{r}\,'')]}\rangle}=\overline{\langle e^{inq_0[u(\vec{r})-u(\vec{0})]}\rangle}.
\end{eqnarray}
Because all of the $ F_n(\vec{r})$ are slowly varying functions of position (because $u(\vec{r}$) itself is), the sum in Eq. (\ref{Xray2}) is dominated by the  term with $nq_0\hat{z}$ the nearest Bragg peak to $\vec{q}$.
Therefore, the X-ray scattering intensity near the $n_{th}$ Bragg peak can be simply written as
\begin{eqnarray}
 I_n(\delta\vec{q})\propto |\rho_n|^2\int d^dr \, e^{i\delta\vec{q}\cdot\vec{r}}F_n(\vec{r}),
 \label{Xray3}
\end{eqnarray}
where $\delta\vec{q}$ is defined as the deviation from the center of the $n_{th}$ Bragg peak, $\delta\vec{q}=\vec{q}-nq_0\hat{z}$.

We have already obtained the scaling of the limiting, large $r$ behavior of $F_n(\vec{r})$ in both the
A and the C phases in section (\ref{Sec: phases}).

To get an idea of the behavior of $F_n(\vec{r})$ for smaller $\vec{r}$'s, which proves to be extremely important for getting the full scattering,  we will begin by approximating  $F_n(\vec{r})$ by
\begin{eqnarray}
 F_n(\vec{r})\thickapprox\exp\left[-{n^2q_0^2\over 2}C(\vec{r})\right],
 \label{Fn}
\end{eqnarray}
where
\begin{eqnarray}
 C(\vec{r})&=& \overline{\langle [u(\vec{r})-u(\vec{0})]^2\rangle}\nonumber\\
           &=&{2\over \left(2\pi\right)^d}\int d^dq\,\left[1-\cos\left(\vec{q}\cdot\vec{r}\right)\right]\overline{\langle u(-\vec{q})u(\vec{q})\rangle}.~~~\nonumber\\
\label{RealuuFunction}
\end{eqnarray}
This expression would be exact if
the fluctuations of the displacement field $u$ were Gaussian. As discussed above, this approximation can be shown\cite{Fisher} to give the correct scaling of $F_n(\vec{r})$ at the longest length scales in both the A and the C phase. Its validity for length scales shorter than the correlation lengths (i.e., for $r_s\ll\xi_s$, $r_z\ll\xi_z$, and $r_h\ll\xi_h$) is more questionable; here we will begin by assuming equation (\ref{RealuuFunction}) and exploring the consequences of that assumption. Then we will discuss possible perils of the approximation (\ref{RealuuFunction}), and argue that the qualitative form of the scattering is given correctly by it, although certain quantitative results (in particular, the precise numerical value of the universal exponent $\Omega$ in equation (\ref{Omegadef}) for the spike width in the X-ray scattering), may be incorrectly given by it.

We can calculate $C(\vec{r})$ in the Gaussian approximation using our knowledge of the disordered averaged fluctuations in momentum space, $\overline{\langle u(-\vec{q})u(\vec{q})\rangle}$.

At large $\vec{q}$s (i.e., $q_s\gg\xi_s^{-1}$, or $q_z\gg\xi_z^{-1}$, or $q_h\gg\xi_h^{-1}$), the term $(\partial_s u)^2$ in the Hamiltonian
Eq. (\ref{CompleteH}) is subdominant to $(\partial_s^2u)^2$
and hence negligible. In this case the random pinning disorder is irrelevant, and we treat the tilt-only model using the RG derived in Sec. \ref{Sec: RG}.
$\overline{\langle u(-\vec{q})u(\vec{q})\rangle}$ can be calculated by following the method described in Sec. \ref{Sec: Anomalous Elasticity}
and given by
\begin{eqnarray}
 \overline{\langle u(-\vec{q})u(\vec{q})\rangle}={\Delta_t(\vec{q}) q_s^2\over \left[B(\vec{q})q_z^2+\gamma q_h^2+K(\vec{q})q_s^4\right]^2}
 \label{u_qc}
\end{eqnarray}
where the wavevector dependences of $B(\vec{q})$, $K(\vec{q})$, and $\Delta_t(\vec{q})$ are given by equations
(\ref{AnomalousB},\ref{AnomalousK},\ref{AnomalousDelta}), respectively.

Plugging $\overline{\langle u(-\vec{q})u(\vec{q})\rangle}$ into Eq. (\ref{RealuuFunction}) we calculate $C(\vec{r})$ asymptotically.
For $\vec{r}$ less than the correlation lengths (i.e., $r_{s,h,z}\ll\xi_{s,h,z}$), the dominant contribution to $C(\vec{r})$ in Eq. (\ref{RealuuFunction}) comes
from the integral over large $\vec{q}\,$'s (i.e., $q_s\gg\xi_s^{-1}$, or $q_z\gg\xi_z^{-1}$, or $q_h\gg\xi_h^{-1}$), and $C(\vec{r})$ is thus given by
\begin{eqnarray}
 C(\vec{r})\approx\lambda_z^2\left\{
 \begin{array}{ll}
 \left(r_s\over\xi_s^N\right)^{\Gamma}, &\left(r_s\over\xi_s^N\right)^{\zeta_z}\gg{r_z\over\xi_z^N}, {r_h\over\xi_h^N}\\
 \left(r_z\over\xi_z^N\right)^{\Gamma/\zeta_z}, &{r_z\over\xi_z^N}\gg\left(r_s\over\xi_s^N\right)^{\zeta_z}, \left(r_h\over\xi_h^N\right)^{\zeta_z/\zeta_h}\\
 \left(r_h\over\xi_h^N\right)^{\Gamma/\zeta_h}, &{r_h\over\xi_h^N}\gg\left(r_s\over\xi_s^N\right)^{\zeta_h}, \left(r_z\over\xi_z^N\right)^{\zeta_h/\zeta_z}
 \end{array}
 \right.\nonumber\\
 \label{SmallCr}
\end{eqnarray}
where $\Gamma=1+\eta_t-\eta_K+{\eta_B\over 2}$.
Plugging this result into Eq. (\ref{Fn}) gives
an exponentially decaying translational correlation function; i.e.,   short-ranged smectic translational  order.

For $\vec{r}$ larger than the correlation lengths (i.e., $r_s\gg\xi_s$, or $r_z\gg\xi_z$, or $r_h\gg\xi_h$),
$C(\vec{r})$ crossovers to
\begin{eqnarray}
  C(\vec{r})\approx &&{1.10\over q_0^2}\ln\left(\sqrt{\left(r_s\over\xi_s\right)^2+\left(r_z\over\xi_z\right)^2+\left(r_h\over\xi_h\right)^2}\right)\nonumber\\
                    &&+\lambda_z^2\left(\xi_s\over\xi_s^N\right)^{\Gamma}.
  \end{eqnarray}
The first piece comes from the integral over small $\vec{q}\,$'s (where we have used ``small" in the sense described earlier). It follows from Eq. (\ref{C_A}) with the cutoff lengths $b_{s,h,z}$ replaced by correlation lengths $\xi_{s,h,z}$, respectively. The second piece comes from the integral over large $\vec{q\,}$'s and connects with $C(\vec{r})$ Eq.(\ref{SmallCr}) at the crossover point.

The x-ray scattering intensity near the $n_{th}$ Bragg peak $I_n(\delta\vec{q})$ can be calculated using Eq. (\ref{Xray3}). Plugging the above calculated $C(\vec{r})$ into Eq. (\ref{Fn}),
we find $F_n(\vec{r})$ has the following behavior. For $\vec{r}$ less than the correlation lengths $F_n(\vec{r})$ decays exponentially fast, while for $\vec{r}$ larger than the correlation lengths $F_n(\vec{r})$ decays much slower and as a power law. Thus, in the large $\delta \vec{q}$ limit the main contribution to $I_n(\delta\vec{q})$ Eq. (\ref{Xray3}) comes from the integral over small $\vec{r}$. This implies that far away from the center of the Bragg peak, the line shape of the scattering intensity is qualitatively Lorentzian-squared. The line widths of this squared-Lorentzian
are anisotropic and given respectively by
\begin{eqnarray}
 \delta q_s^w&=&\left(\xi_s^w\right)^{-1}\approx \left(\xi_s^N\right)^{-1}\left(a\over\lambda_z\right)^{-2/\Gamma},\\
 \delta q_{z,h}&=&\left(\xi_{z,h}^w\right)^{-1}\approx \left(\xi_{z,h}^N\right)^{-1}\left(a\over\lambda_z\right)^{-2\zeta_{z,h}/\Gamma},~~~~~~
\end{eqnarray}
where $\xi_{s,z,h}^w$ are the line widths of the correlation function $F_n(\vec{r})$ and obtained via $C(\vec{r})=a^2$. In the limit $\delta \vec{q}\to \vec{0}$, $I_n(\delta\vec{q})$ converges for $n>2$ and diverges for $n\leq 2$. Therefore, only the first two Bragg peaks diverge, following the power law:
\begin{eqnarray}
 I_n(\delta\vec{q})\propto \varpi_n\left[\left(\delta q_s\xi_s\right)^2+\left(\delta q_h\xi_h\right)^2+\left(\delta q_z\xi_z\right)^2\right]^{-3+.55n^2\over 2}\nonumber\\
 \label{SmallXray}
\end{eqnarray}
where $\varpi_n=\xi_s\xi_z\xi_h\exp\left[-{1\over 2}n^2\lambda_z^2q_0^2\left(\xi_s/\xi_s^N\right)^{\Gamma}\right]$ has a exponentially strong dependence on the temperature, which is responsible for the fast vanishing of the divergent Bragg peak approaching the critical point. The divergence shown by (\ref{SmallXray}) comes from the integral in Eq. (\ref{Xray3}) over large $\vec{r}$.

The crossover between these two distinct scattering patterns can be estimated as the follows. Since the divergent (power law) Bragg peak vanishes extremely fast approaching the critical point, we expect that near the critical point, the crossover occurs at very small $\delta\vec{q}$s which are much less than the widths of the Lorentzian square. Thus, at this crossover the Lorentzian part of $I_n(\delta\vec{q}_c)$ is approximately proportional to $\xi_s^w\xi_h^w\xi_z^w$. Equating this to the power law part given by Eq. (\ref{SmallXray}) we get, for the $n_{th}$ Bragg peak,
\begin{eqnarray}
 \delta q_{s,h,z}^{nc}&=&\xi_{s,h,z}^{-1}\left(\xi_s^w\xi_h^w\xi_z^w\over\varpi_n\right)^{1\over -3+.55n^2}\nonumber\\
                      &\propto& t^{\Upsilon_{s,h,z}^n}\exp\left(-A_nt^{\Omega_G}\right),
                      \label{XrayCrossoverGaussian}
\end{eqnarray}
where $A_n$ is non-universal, $\Omega_{G}$ (where the subscript ``$G$" stands for ``Gaussian") and $\Upsilon_{s,h,z}^n$ are universal and given, respectively, by
\begin{eqnarray}
 \Omega_G&=&\nu_s\Gamma={1\over 2}+{4\over 27}\epsilon+O(\epsilon^2),
 \label{Omega Gaussian}
 \end{eqnarray}
 \begin{eqnarray}
 \Upsilon_{s,h,z}^n&=&\nu_{s,h,z}+{\nu_s\nu_h\nu_z\over -3+.55n^2}.
 \label{Upsilon}
\end{eqnarray}

These results were derived in the Gaussian approximation. Since we have shown that the smectic elasticity
and fluctuations are controlled by a non-Gaussian fixed point, one might well question the validity of the Gaussian approximation.

How could we go beyond the Gaussian approximation? In principle, this could be done by integrating all of the degrees of freedom out of the  Hamiltonian ({completeH}) except for the single variable that enters the X-ray scattering near the $n$'th Bragg peak, namely:
\begin{eqnarray}
x_n(\vec{r})\equiv nq_0[u(\vec{r})-u(\vec{0})]~.
                      \label{xdef}
\end{eqnarray}
The hypothetical integration out of all other variables would leave us with a known probability distribution $P(x_n(\vec{r}))$ of this single variable $x_n(\vec{r})$.

In reality, of course, it is far too difficult to perform this hypothetical integration. We do, however, know {\it something} about $P(x_n(\vec{r}))$: its variance, which is just $C(\vec{r})$,  equation (\ref{RealuuFunction}). This sets the scale of  $x_n(\vec{r})$ fluctuations; hence, we can write $P(x_n(\vec{r}))$ in a scaling form:
\begin{eqnarray}
P(x_n(\vec{r}))=\sqrt{k}e^{-U(\sqrt{k(\vec{r})} x_n(\vec{r}))}~,
                      \label{xdist}
\end{eqnarray}
where
\begin{eqnarray}
k(\vec{r})\equiv {O(1)\over (n q_0)^2 C(\vec{r})}~.
\label{k def}
\end{eqnarray}

If we take $U(y)=y^2$, then we recover the Gaussian theory just presented. More generally, we can write:
\begin{eqnarray}
F_n(\vec{r})= \sqrt{k}\int^{\infty}_{-\infty} dx ~e^{i x-U(\sqrt{k(\vec{r})} x)} ~.
                      \label{Fn gen}
\end{eqnarray}

For $|\vec{r}|$ large, $k(\vec{r})$ is small, and one can accurately evaluate this integral, for a given $U(y)$,  using the method of saddle points. That is, one finds the place where the argument of the exponential in (\ref{Fn gen}) is extremized, expands about that point, and performs a Gaussian integral using this expansion. In general, locating the saddle point  requires analytically continuing the integral in (\ref{Fn gen}) into the complex plane.

The first step is, of course, locating the saddle point.
Defining
\begin{eqnarray}
f(x)\equiv ix-U(\sqrt{k(\vec{r})} x)~,
  \label{f def}
\end{eqnarray}
this point is where
\begin{eqnarray}
f'(x)= i-\sqrt{k(\vec{r})} U'(\sqrt{k(\vec{r})} x)=0~.
  \label{saddle cond}
\end{eqnarray}

\begin{figure}
 \includegraphics[width=0.4\textwidth]{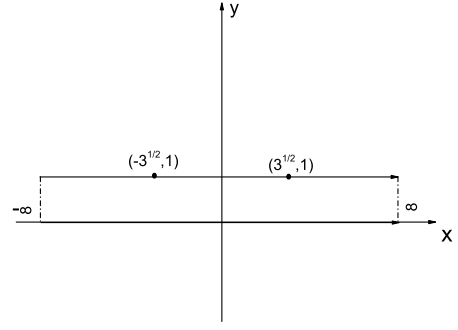}
 \caption{\label{fig: contour}The complex contour we choose for the saddle point evaluation of the integral (\ref{Fn gen}) for the translational correlation function $F_n(\vec{r})$.}
\end{figure}

For example, consider $U(y)=y^4$\cite{O(1) comment}.
The condition (\ref{saddle cond}) becomes
\begin{eqnarray}
4 k^2 x^3=i~,
  \label{saddle cond quartic}
\end{eqnarray}
whose solution is \cite{saddle}:
\begin{eqnarray}
x_\pm = \left({i\over 4k^2}\right)^{1\over 3}={(\pm\sqrt{3}+i)\over{(32 k^2)^{1\over 3}}}~. \label{saddle soln}
\end{eqnarray}
To exploit these saddle points, we deform our contour of integration from the real axis to a line parallel to the real axis, but displaced up (i.e., in the imaginary direction) from it by the imaginary part of $x_\pm$, as illustrated  in Fig. \ref{fig: contour}. (The contour is connected to the original contour by vertical segments at $\pm\infty$ that contribute nothing to the integral.) This passes the contour of integration through both saddle points $x_\pm$.

Since the integral is now dominated by the region near these two saddle points (as we will verify a posteriori in a moment), we can expand the integrand around each of the saddle points. Noting that
\begin{eqnarray}
f''(x_\pm) =-12k^2x_\pm^2=-6  \left({k\over 4}\right)^{2\over 3}(1\pm\sqrt{3}i)
\label{saddle curve}
\end{eqnarray}
and
\begin{eqnarray}
f(x_\pm) =i x_\pm-k^2x_\pm^4={3(\pm\sqrt{3}i-1)\over8(2 k)^{2\over 3}}~,
\label{saddle f}
\end{eqnarray}
we can use the expansion
\begin{eqnarray}
f(x) \approx f(x_\pm) +{1\over 2}  f''(x_\pm) u^2
\label{saddle approx}
\end{eqnarray}
with $u$ real, near each of the saddle points. Making this approximation, and extending the region of integration over $u$ near each saddle point to run from $-\infty$ to $\infty$ (an approximation that we will justify for small $k$ in a moment), we are then left with two Gaussian integrals, whose sum is $F_n(\vec{r})$. This gives 
\begin{eqnarray}
&&F_n(\vec{r})\approx\sqrt{{2\pi\over 3}} \left({4\sqrt{k}}\right)^{1\over 3} {\rm exp}\left(-{3\over 8(2 k)^{2\over 3}}\right) \nonumber\\&&  \times{\rm cos}\left({3\sqrt{3}\over 8(2 k)^{2\over 3}}-{\pi\over 6}\right)
            \label{Fn quartic}
\end{eqnarray}

Now, to justify the saddle point approximation:

This approximation will be valid provided that the expansion (\ref{saddle approx})
is valid over the range of $u$ that dominates the integral near each of the saddle points. This implies that it must work for
\begin{eqnarray}
u\sim{1\over\sqrt{f''(x_\pm)}}~.
\label{int width}
\end{eqnarray}
The neglected next order term in the expansion (\ref{saddle approx}) is $\sim f'''(x_\pm) u^3\sim { f'''(x_\pm) \over f''(x_\pm)^{3\over 2}}$, where we have used the estimate (\ref{int width})
for the range of $u$ that dominates the integral near each saddle point. As long as this is $\ll1$, the saddle point approximation will be good. Since, from (\ref{saddle curve}), $f''(x_\pm)\sim k^{2\over 3}$   while $f'''(x_\pm)\sim k^2x_\pm \sim k^2 k^{-{2\over 3}}\sim k^{4\over 3}$ (where we have used (\ref{saddle soln}) for $x_\pm$), we have
${ f'''(x_\pm) \over f''(x_\pm)^{3\over 2}}\sim{k^{4\over 3}\over k}\sim k^{1\over 3} \ll 1$, provided $k\ll1$, as it is for large $|\vec{r}|$, as noted earlier.

Thus, for large $|\vec{r}|$, the saddle point approximation is valid.

More generally, if we take $U(y)=y^n$, then we get, by exactly the same sort of steepest descent analysis,

\begin{eqnarray}
&&F_n(\vec{r})\propto{\rm exp}\left(-O(1) \times k^{-\Theta(n)}\right)
            \label{Fn gen power}
\end{eqnarray}
with
\begin{eqnarray}
\Theta(n)\equiv {n\over2(n-1)}~.
       \label{Theta def}
\end{eqnarray}

This is readily seen to reproduce the results for the Gaussian case ($n=2$), and the quartic case we just examined in detail ($n=4$). Note that for all values of $n\ge2$,
the exponent $\Theta$ is bounded above and below:
\begin{eqnarray}
{1\over 2}\le\Theta(n)\le 1~,
       \label{Theta bound}
\end{eqnarray}
with the upper bound satisfied by the Gaussian case, while the lower bound is approached in the limit $n\rightarrow \infty$. Since we expect $n\ge 2$, since H contains higher powers of u and the value of $\Theta$ is determined by the highest power, we expect $\Theta$ to satisfy (\ref{Theta bound}).

As a final example, let us consider $U(y)=e^{-y}$.
The steepest descent analysis now gives
\begin{eqnarray}
&&F_n(\vec{r})\propto{\rm exp}\left(-O(lnk) \times k^{-{1\over 2}}\right) ~,
            \label{Fn exp}
\end{eqnarray}
which is effectively (up to slowly varying logarithmic corrections) of the form (\ref{Fn gen power})
with $\Theta={1\over 2}$; i.e., just the limiting value of $\Theta(n)$ as $n\rightarrow \infty$ (as is hardly surprising, since the Taylor series for $e^{-y}$
involves arbitrarily large powers of $y$).

While hardly exhaustive, these examples give us confidence that the leading behavior of $F_n(\vec{r})$ is always given by (\ref{Fn gen power}), with $\Theta$ satisfying (\ref{Theta bound}). Using this, and the definition (\ref{k def}), and repeating the analysis  given above of  the Gaussian case, we find that $ \delta q_{s,h,z}^{nc}$ is given by:
\begin{eqnarray}
 \delta q_{s,h,z}^{nc}&=&\xi_{s,h,z}^{-1}\left(\xi_s^w\xi_h^w\xi_z^w\over\varpi_n\right)^{1\over -3+.55n^2}\nonumber\\
                      &\propto& t^{\Upsilon_{s,h,z}^n}\exp\left(-A_nt^{\Omega}\right),
                      \label{XrayCrossover}
\end{eqnarray}
with
\begin{eqnarray}
\Omega=\Omega_G\Theta~,
                      \label{Omega true}
\end{eqnarray}
where $\Omega_G$ is given by (\ref{Omega Gaussian}), $\Upsilon$ by (\ref{Upsilon}), and $\Theta$ obeys the bound (\ref{Theta bound}). Combining this bound on $\Theta$ with an estimate of the uncertainties in the $\epsilon$-expansion result (\ref{Omega Gaussian}) for $\Omega_G$ leads to the bounds on $\Omega$ quoted in the introduction.

Near the critical point the dominant $t$-dependence of  $\delta q_{s,h,z}^{nc}$ is exponential. Thus, a plot of $\ln{\ln{\delta q_{s,h,z}^{nc}}}$
versus $\ln{t}$ should give a straight line with slope $\Omega$.

\subsection{\label{Sec: light scat}  Visible Light Scattering}

The fluctuations of the nematic director scatter the light strongly. The scattering intensity is proportional to the fluctuations \cite{Lubensky}:
\begin{eqnarray}
 I(\vec{q})&\propto& \left(\hat{e}_{\ell}\cdot\vec{n}_0\right)^2\overline{\langle |\delta\vec{n}(\vec{q})\cdot\hat{e}_T|^2\rangle}+\left(\hat{e}_T\cdot\hat{n}_0\right)^2\overline{\langle |\delta\vec{n}(\vec{q})\cdot\hat{e}_{\ell}|^2\rangle}\nonumber\\
 &&+2\left(\hat{e}_{\ell}\cdot\hat{n}_0\right)\left(\hat{e}_T\cdot\hat{n}_0\right)\hat{e}_{\ell}\cdot\overline{\langle \delta\vec{n}(\vec{q})\delta\vec{n}(-\vec{q})\rangle}\cdot\hat{e}_T,\nonumber\\
\end{eqnarray}
where $\hat{e}_{\ell}$ and $\hat{e}_T$ are respectively the polarization direction of the incident and the transmitted light, and $\delta\hat{n}$ is the deviation of the nematic director from its averaged value $\hat{n}_0$: $\delta\vec{n}\equiv\hat{n}-\hat{n}_0$.

In the $A$ phase $\hat{n}_0=\hat{z}$, so the  largest components of $\delta\vec{n}$ are those  $\delta \vec{n}_{\perp}$ perpendicular to $z$. Thus, the scattering intensity is proportional to a linear sum of $\overline{\langle\delta n^{\perp}_i\delta n^{\perp}_j\rangle}$; namely,
\begin{eqnarray}
I(\vec{q})\sim\sum_{ij}D_{ij}\overline{\langle
\delta n_i^{\perp}(\vec{q})\delta n_j^{\perp}(-\vec{q})\rangle},\label{Scattern}
\end{eqnarray}
where the coefficients $D_{ij}$ are of order one and solely determined by the polarization directions of the incident and transmitted light, and $\delta n_i^{\perp}$ denotes the $i\,$'th component of $\delta \vec{n}_{\perp}$.

Since, in Sec. \ref{Sec: Model}, we have shown that the fluctuations of the   linear combination $\vec{\delta n'}_{\perp}$ (equation (\ref{massive}))  of $\delta n_i^{\perp}$ and $\partial_i^{\perp}u$ are massive, $\overline{\langle
\delta n_i^{\perp}(-\vec{q})\delta n_j^{\perp}(\vec{q})\rangle}\approx \overline{\langle\partial_i^{\perp}u(-\vec{q})\partial_j^{\perp}u(\vec{q})\rangle}$,
where $\partial^{\perp}_i$ denotes the partial derivative along the $i\,$'th component in the $\perp$ plane. Therefore, equation (\ref{Scattern}) can be rewritten as
\begin{eqnarray}
I(\vec{q})\sim\sum_{ij}E_{ij}C_{ij}(\vec{q})\label{LightScattering}
\end{eqnarray}
where
\begin{eqnarray}
 C_{ij}(\vec{q})\equiv L_{ij}(\hat{q})q_{\perp}^2\overline{\langle |u(\vec{q})|^2\rangle}~,
 \label{C light}
\end{eqnarray}
with
$L_{ij}(\hat{q})\equiv q_i^{\perp}q_j^{\perp}/q_{\perp}^2$  the projection operator along $\perp$, and the coefficients $E_{ij}$
are independent of $\vec{q}$ and of order one.

An expression for  $\overline{\langle |u(\vec{q})|^2\rangle}$ has been obtained in Sec. \ref{Sec: X-ray}.
However, those results have to be improved to predict the light scattering. Specifically,
in addition to the fluctuations induced by the random tilt in the $\hat{s}$ direction and the random field,
we need to include in $\overline{\langle |u(\vec{q})|^2\rangle}$ the fluctuations induced by the random tilt in
$\hat{h}$ direction and the random compression, as embodied in the $\Delta_h$ and $\Delta_c$ terms of equation (\ref{Hm}). These last two types of disorder, though `` irrelevant" in the RG sense, have experimentally detectable effects  on light scattering.

If {\it any} component of   $\vec{q\,}$ is  large compared to the corresponding inverse correlation length, (i.e., if $q_s\gg \xi_s^{-1}$, {\it or} $q_h\gg \xi_h^{-1}$, {\it or}
$q_z\gg \xi_z^{-1}$), but all components are small compared to the ultraviolet cutoff $\Lambda$,
we find that the $\overline{<|u(\vec{q})|^2>}$ in (\ref{C light}) is given to a good approximation by
\begin{eqnarray}
 \overline{<|u(\vec{q})|^2>}&=&
{\Delta_t\left(\vec{q}\right)q_s^2 +\Delta_hq_h^2+
 \Delta_c\left(\vec{q}\right)q_z^2
\over \left(B(\vec{q})q^2_z
 +\gamma q_h^2+K\left(\vec{q}\right)q^4_s\right)^2}
~, \nonumber\\
\label{CfunctionCrit}
\end{eqnarray}
where the $\vec{q}$-dependences of $B(\vec{q})$, $K(\vec{q})$, and $\Delta_{t}(\vec{q})$ were derived in section \ref{Sec: Anomalous Elasticity}, and are given
by Eqs. (\ref{AnomalousB}, \ref{AnomalousK}, and \ref{AnomalousDelta}, respectively. The wavevector dependence $\Delta_{c}(\vec{q})$ of the $z$ component of the random field $\vec{h}$  is derived in Appendix B,
and given by equation (\ref{AnomalousDelta_c}). The correlations $\Delta_{h}(\vec{q})$ of the $h$ component of the random field $\vec{h}$ do {\it not} acquire any anomalous wavevector dependence
since there is no graphical correction to $\Delta_{h}(\vec{q})$.
The fluctuations caused by the random field disorder are not included in Eq. (\ref{CfunctionCrit}),  since their contributions
to $\overline{<|u(\vec{q})|^2>}$ are always dominated by the others for large $\vec{q}$.

Likewise, if {\it all} of the components of   $\vec{q}\,$ are small compared to the corresponding  inverse correlation length, (i.e.,  if $q_s\ll \xi_s^{-1}$, {\it and} $q_h\ll \xi_h^{-1}$, {\it and}
$q_z\ll \xi_z^{-1}$), we obtain
\begin{widetext}
\begin{eqnarray}
\overline{<|u(\vec{q})|^2>}= {\Delta_t \left( T
\right)q_s^2 +\Delta_hq_h^2+\Delta_c\left( T \right)q_z^2\over \left(B(T)q^2_z+\gamma q_h^2+D(T)q^2_s\right)^2}+{1.10\pi^2\sqrt{B(T)D(T)\gamma}\over\left(B(T)q^2_z+\gamma q_h^2+D(T)q^2_s\right)^{3\over 2}q_0^2}~,
\label{CfunctionA}
\end{eqnarray}
\end{widetext}
where the third piece comes from the random field disorder.
Recall that
 $B(T)$, $K(T)$, $D(T)$, $\Delta_{t, c}(T)$ are no longer wavevector dependent in this regime of wavevector, , but {\it are} renormalized by the critical fluctuations,
and, hence, temperature-dependent; this temperature dependence is summarized in equations  \ref{critB} \ref{critK}, \ref{critD}, \ref{critDt}, \ref{critDc}.

As mentioned in the introduction, These rather complicated expressions (equations  (\ref{CfunctionCrit}) and  (\ref{CfunctionA})) simplify considerably in a variety of natural limits.

Consider first varying wavenumber $q$ for fixed, generic {\it direction} $\hat{q}$ of the scattering wavevector $\vec{q}$.  By ``generic direction $\hat{q}$ ", we mean a direction of $\hat{q}$ for which  no component of  $\hat{q}$ is $\ll 1$.

Consider first the regime of  $\vec{q\,}$'s large compared to the correlation lengths, (i.e., $q_s\gg \xi_s^{-1}$, or $q_h\gg \xi_h^{-1}$, or
$q_z\gg \xi_z^{-1}$), but small compared to the ultraviolet cutoff $\Lambda$, in which equation (\ref{CfunctionCrit})  is valid. Since all three components of $\vec{q}$ are comparable for a generic direction of $\hat{q}$,  this amounts to saying that the magnitude $q$ of $\vec{q}$ must be large compared to the {\it smallest} of the {\it inverse} correlation lengths $\xi_s^{-1}$,  $q_h\gg \xi_h^{-1}$, and
$q_z\gg \xi_z^{-1}$, which means it must be large compared to the inverse of
the {\it biggest} correlation length, which is $\xi_h$, since $\xi_h\propto\xi_s^{\zeta_h}\gg\xi_z\propto\xi_s^{\zeta_z}$, where the inequality follows from the facts that $\zeta_h=2-{\eta_K\over 2} > \zeta_z=2-{\left(\eta_K+\eta_B\right)\over 2}$ and $\xi_s\rightarrow\infty$ as the transition is approached.

Thus, we are, for a generic direction of $\hat{q}$, in the regime in which equation (\ref{CfunctionCrit}) is valid, provided
\begin{eqnarray}
q\gg\xi_h^{-1}\propto|T-T_{AC}|^{\nu_h}~~,
\label{generic q crossover}
\end{eqnarray}
where $\nu_h$ is given by equation (\ref{nu_eps}).

For $q\,$'s satisfying the condition (\ref{generic q crossover}), with a generic
direction of $\hat{q}$,  the dominant cutoff in all of the scaling functions for all of the anomalous elastic coefficients in equation (\ref{CfunctionCrit}) is $q_h$.

 To see this, begin by noting that the scaling arguments
 \begin{eqnarray}
X\equiv {q_z\xi_z^N\over \left(q_s\xi_s^N\right)^{\zeta_z}}~~,
\end{eqnarray}
and
\begin{eqnarray}
Y\equiv {q_h\xi_h^N\over \left(q_s\xi_s^N\right)^{\zeta_h}}~~.
\end{eqnarray}
that appear in all of the scaling functions  (e.g., equation (\ref{ell*}) for $\ell^*$) are both $\gg 1$, since $X\propto q^{1-\zeta_z}\rightarrow \infty$ and $
Y\propto q^{1-\zeta_h}\rightarrow \infty$ as  $q\rightarrow \infty$, since both $\zeta_z$ and $\zeta_h$ are $> 1$. This implies that either $q_h$ or $q_z$ dominates the scaling functions. To determine which, we form the only possible
$q_s$-independent ratio of $X$ and a power of $Y$: ${X\over Y^{\zeta_z\over\zeta_h}}={q_z\xi_z^N\over{\left(q_h\xi_h^N
\right)^{\zeta_z\over \zeta_h}}}\propto q^{1-{\zeta_z\over \zeta_h}}$
$=q^{{\eta_B \over 4-\eta_K}}\rightarrow 0$ as  $q\rightarrow
\infty$. The  vanishing  of this ratio implies that $q_h$ is the dominant cutoff in all of the scaling functions.

Using this fact implies that, for the generic directions
of $\vec{q}$ that we are considering, $\Delta_t(\vec{q})\propto q^{-{\eta_t\over\zeta_h}}$, and
$\Delta_c(\vec{q})\propto q^{-{\eta_c\over\zeta_h}}$. This in turn
implies that
the $\Delta_c(\vec{q})$ term in the numerator of eqn. (\ref{CfunctionCrit}) dominates the $\Delta_t(\vec{q})$ and $\Delta_h$
terms, since the ratio of the $\Delta_t(\vec{q})$ term to the $\Delta_c(\vec{q})$ term vanishes like $q^{\eta_c-\eta_t \over
\zeta_h}$, which vanishes as $q\rightarrow 0$, since $
\eta_c-\eta_t=2-\eta_B-\eta_K > 0$; likewise, the ratio of the $\Delta_h$ term to the
$\Delta_c(\vec{q})$ term vanishes like $q^{\eta_c\over\zeta_h}$, which vanishes as $q\rightarrow 0$, since $
\eta_c> 0$.

Similar arguments show that the $\gamma$ term, which scales like $q^2$   for the generic directions
of $\vec{q}$ that we are considering, dominates the denominator of eqn. (\ref{CfunctionCrit}).

Taking these two terms as dominant in the numerator and denominator of eqn. (\ref{CfunctionCrit}), respectively, we find:
\begin{eqnarray}
\overline{<|u(\vec{q})|^2>}\propto q^{-2-{\eta_c\over\zeta_h}}{\hat{q}_z^2\over\hat{q}_h^{4+{\eta_c\over\zeta_h}}}~~,  ~~q\gg\xi_h^{-1}~,
\label{CgenericCrit}
\end{eqnarray}
for generic directions of $\hat{q}$. Using this result in equation (\ref{C light}), we obtain for the light scattering intensity, in this regime of wavevector:
\begin{eqnarray}
I(\vec{q})\propto q^{-{\eta_c\over\zeta_h}}{\hat{q}_z^2\over\hat{q}_h^{4+{\eta_c\over\zeta_h}}}~~, ~~q\gg\xi_h^{-1}~.
\end{eqnarray}
Using the epsilon-expansion results (\ref{epsDc}) and (\ref{epszetah}) for $\eta_c$ and $\zeta_h$, we obtain for the $\epsilon$-expansion for the power law in this expression:
\begin{eqnarray}
{\eta_c\over\zeta_h}=1-{8\epsilon\over 27}+O(\epsilon^2)~~,
\label{light exp}
\end{eqnarray}

As discussed above, once $q_h$ decreases to
$\xi_h^{-1}$,  the correlation function $<|u(\vec{q})|^2>$ crosses over to the form equation(\ref{CfunctionA}). In this regime, as mentioned earlier, all of the elastic moduli and disorder variances cease to be wavevector dependent, but become temperature dependent, as given by equations (\ref{critB}-\ref{critDc}). As $q$ first decreases below $\xi_h^{-1}$, the $\Delta_c$ term continues to dominate $<|u(\vec{q})|^2>$; in this regime, this implies $<|u(\vec{q})|^2>\propto{1\over q^2}$, and, hence, that  the light scattering intensity $I(\vec{q})$ is constant. Eventually, however, as $q$ continues to decrease, the ``random field'' term (i.e., the $\sqrt{DB\gamma}$ term, must start to dominate, since it diverges like ${1\over q^3}$ as $q\rightarrow 0$. For smaller $q\,$'s,  $<|u(\vec{q})|^2>\propto{1\over q^3}$, and, hence, that  the light scattering intensity $I(\vec{q})$ obeys
\begin{eqnarray}
I(\vec{q})\propto q^{-1}~.
\end{eqnarray}
We can estimate the critical value $q^F$ of $q$ at which this crossover between constant $I(\vec{q})$ and $I(\vec{q})\propto q^{-1}$takes place by equating the $\Delta_c$ term and the random field term; i.e., by setting:
\begin{eqnarray}
{\Delta_c\left( T \right)(q^F_z)^2\over\gamma^2 (q^F_h)^4}={\sqrt{B(T)D(T)\gamma}\over\left(\gamma (q^F_h)^2\right)^{3\over 2}q_0^2}~,
\label{q_c cond}
\end{eqnarray}
where we have used the fact that the $q_h$ term continues to dominate the denominators of all terms for this generic direction of $\vec{q}$, since its coefficient $\gamma$ is the only one in the denominator that is not renormalized to small values by the critical fluctuations.
Solving equation (\ref{q_c cond}) for $q_c$ gives:
\begin{eqnarray}
q^F(\hat{q})={\sqrt{B(T)D(T)}\gamma\over\left(\Delta_c\left( T \right) \hat{q}_h^2 \hat{q}_z^2\right)q_0^2}\propto |T-T_{AC}|^{\phi}~,
\label{q_c sol}
\end{eqnarray}
where the universal exponent $\phi$ is given by
\begin{eqnarray}
\phi=\nu_s\left(1+\eta_c+{\eta_B-\eta_K\over 2}\right)={3\over 2}+{8\over 27}\epsilon+O(\epsilon^2)~,\nonumber\\
\label{phi2}
\end{eqnarray}
where the first equality is obtained using the known temperature dependences equations (\ref{critB}), (\ref{critD}), and (\ref{critDc}) for $B(T)$, $D(T)$, and $\Delta_c(T)$, while the second follows from the known $\epsilon$-expansions (\ref{epsK}),(\ref{epsB}), and (\ref{epsDc}),  for $\eta_K$, $\eta_B$, and $\eta_c$. Equation (\ref{phi2}) is, of course, precisely equation (\ref{phi})
of the Introduction.

To summarize, for generic directions $\hat{q}$ of $\vec{q}$, the light scattering is given by equation (\ref{Iqgen}) of the Introduction, and plotted there in Fig. \ref{fig: lightgen}.

As mentioned in the introduction, more information can be obtained by restricting the scattering to the subspace of $\vec{q}$ with $q_h=0$.  In this case the
only nonzero $C_{ij}(\vec{q})$ is $C_{ss}(\vec{q})$. Therefore, equation (\ref{LightScattering}) implies that
\begin{eqnarray}
I(\vec{q})\propto
C_{ss}=q_s^2\overline{\langle |u(\vec{q})|^2\rangle}~~.
\label{I_qh=0}
\end{eqnarray}

As we just did for the case of a generic direction of $\vec{q}$, here, too, we can distinguish between the cases of  $\vec{q\,}$'s large compared to the correlation lengths, (which  now means $q_s\gg \xi_s^{-1}$, or
$q_z\gg \xi_z^{-1}$), but small compared to the ultraviolet cutoff $\Lambda$, and the reverse.
In the large $\vec{q}$ case,
the correlation function $\overline{\langle |u(\vec{q})|^2\rangle}$  takes a simple scaling form:
\begin{eqnarray}
\overline{\langle |u(\vec{q})|^2\rangle}=f_u(X)q_s^{2\eta_k-6-\eta_t}~~,
\label{uscale}
\end{eqnarray}
where $X$ is the scaling ratio defined in equation (\ref{Xdef}).
This can be seen by noting that, when $q_h=0$, both the numerator and denominator of (\ref{CfunctionCrit}) take a scaling form. The numerator is
\begin{eqnarray}
\Delta_t\left(\vec{q}\right)q_s^2 &+&
 \Delta_c\left(\vec{q}\right)q_z^2=q_s^{2-\eta_t}f_t(X)+q_s^{-\eta_c}q_z^2f_c(X) \nonumber \\
 &=&q_s^{2-\eta_t}\left(f_t(X)+q_z^2q_s^{\eta_t-2-\eta_c}f_c(X)\right) ~~,\nonumber \\
 \label{numscale1}
\end{eqnarray}
where we have used the scaling forms (\ref{AnomalousDelta}) and (\ref{AnomalousDelta_c}) for $\Delta_t$ and $\Delta_c$.

Using the exact scaling relation (\ref{ExactEta_c}) for $\eta_c$, we find that $\eta_t-2-\eta_c=-4+\eta_B+\eta_K=-2\left(2-{\eta_B+\eta_K\over 2}\right)=-2\zeta_z$, where we have also used the exact scaling law $\zeta_z=2-{\eta_B+\eta_K\over 2}$.

Using this in (\ref{numscale1}) implies that
\begin{eqnarray}
\Delta_t\left(\vec{q}\right)q_s^2 &+&
 \Delta_c\left(\vec{q}\right)q_z^2=q_s^{2-\eta_t}\left(f_t(X)+\left({q_z\over q_s^{\zeta_z}}\right)^2f_c(X)\right) \nonumber \\
 &\equiv&q_s^{2-\eta_t}f_{\rm{num}}(X)~~,\nonumber \\
 \label{numscale2}
\end{eqnarray}
where we have defined the scaling function for the numerator
\begin{eqnarray}
f_{\rm{num}}(X)\equiv f_t(X)+\left[{\left(\xi_s^N\right)^{\zeta_z}\over \xi_z^N}\right]^2X^2f_c(X)~~,
\label{fnumdef}
\end{eqnarray}
and used the definition (\ref{Xdef}) of the scaling variable $X$.

Similarly, the square root of the denominator of (\ref{CfunctionCrit}) can be written, when $q_h=0$,  as:
\begin{eqnarray}
B\left(\vec{q}\right)q_z^2 &+&
K\left(\vec{q}\right)q_s^4\nonumber \\
&=&q_s^{4-\eta_K}\left({q_z^2\over q_s^{4-\eta_B-\eta_K}}f_B(X)+f_K(X) \right) ~~,\nonumber \\
&\equiv&q_s^{4-\eta_K}f_{\rm{den}}(X)
 \label{denscale1}
\end{eqnarray}
where we have  used used the scaling forms (\ref{AnomalousB}) and (\ref{AnomalousK}) for $B$ and $K$,  we have defined the scaling function for the denominator
\begin{eqnarray}
f_{\rm{den}}(X)\equiv f_K(X)+\left[{\left(\xi_s^N\right)^{\zeta_z}\over \xi_z^N}\right]^2X^2f_B(X)~~,
\label{fdendef}
\end{eqnarray}
and we have once again used both the definition (\ref{Xdef}) of the scaling variable $X$, and the exact scaling relation $2-{\eta_B+\eta_K\over 2}=\zeta_z$.

Using the final equalities of  (\ref{numscale1}) and  (\ref{denscale1}) in (\ref{CfunctionCrit}) implies the scaling form equation (\ref{uscale}), with
\begin{eqnarray}
f_u(X)\equiv {f_{\rm{num}}(X)\over f^2_{\rm{den}}(X)}~~.
\label{fudef}
\end{eqnarray}
and used the definition (\ref{Xdef}) of the scaling variable $X$.

The scaling function $f_u(X)$ must have limits such that the full $\overline{\langle |u(\vec{q})|^2\rangle}$ correlation function becomes a function only of $q_s$ for $X\ll 1$, and only of $q_z$ for $X\gg 1$. This implies, by arguments virtually identical to those  used in, e.g., section(\ref{Sec: Anomalous Elasticity}),
\begin{eqnarray}
 f_u(X)\propto\left\{\begin{array}{ll}
 \rm{constant}, ~~&X\ll 1~, \\
 X^{{2\eta_k-6-\eta_t\over\zeta_z}}, ~~&X\gg 1~,
 \end{array}
 \right.
\end{eqnarray}
which in turn implies
\begin{eqnarray}
\overline{\langle |u(\vec{q})|^2\rangle}\propto\left\{\begin{array}{ll}
q_s^{2\eta_k-6-\eta_t}, ~~&q_z\xi_z^N \ll
\left(q_s\xi_s^N\right)^{\zeta_z},
 \\
 q_z^{{2\eta_k-6-\eta_t\over\zeta_z}}, ~~&q_z\xi_z^N \gg
\left(q_s\xi_s^N\right)^{\zeta_z}~.
 \end{array}
 \right.
 \label{ulsqh=0crit}
\end{eqnarray}

Using equation (\ref{I_qh=0}), we can obtain the light scattering intensity from this:
\begin{eqnarray}
I(\vec{q})\propto\left\{\begin{array}{ll}
q_s^{2\eta_k-4-\eta_t}, ~~&q_z\xi_z^N \ll
\left(q_s\xi_s^N\right)^{\zeta_z},
 \\
q_s^2 q_z^{{2\eta_k-6-\eta_t\over\zeta_z}}, ~~&q_z\xi_z^N \gg
\left(q_s\xi_s^N\right)^{\zeta_z}~.
 \end{array}
 \right.
 \label{Iqh=0crit}
\end{eqnarray}

The above result only applies for both $q_{s,z}$ each much larger than its associated inverse
correlation length $\xi^{-1}_{s,z}$. Once this ceases to be true, there will, as in the case of a generic direction of $\vec{q}$ just discussed, be two subsequent crossovers   as we decrease $\vec{q}$: first, at $q_{s,z}=\xi^{-1}_{s,z}$, at which point  all of the parameters $\Delta_{t,c}$, $B$, and $K$ cease to be wavevector dependent; (i.e., when the crossover between Eqs. (\ref{CfunctionCrit}) and (\ref{CfunctionA}) occurs). Then, as we continue lowering $q$, at a smaller $q$, a crossover from
the $u-u$ correlations being dominated by random compressions and tilts to being dominated by the random field term (i.e., crossover from the first to the second term dominating in equation (\ref{CfunctionA}).
Taking all this into account leads to different behaviors of $\overline{\langle |u(\vec{q})|^2\rangle}$ in the six different regions of the
 $q_h$-$q_z$, $q_h=0$ plane illustrated in Fig. \ref{fig: lightsz}.

The locus $FG$, given by
\begin{eqnarray}
FG:~~q_z\xi_z^N =
\left(q_s\xi_s^N\right)^{\zeta_z}~,
\label{FGdef}
\end{eqnarray}
is where the crossover between the two limits of equation (\ref{Iqh=0crit}) occurs.

The locus $EF$, given by
\begin{eqnarray}
EF:~~q_z =\xi_z^{-1},
\label{EFdef}
\end{eqnarray}
is where we crossover between expressions
Eqs. (\ref{CfunctionCrit}) and (\ref{CfunctionA}) on the
$q_z\xi_z^N \gg
\left(q_s\xi_s^N\right)^{\zeta_z}$ side of $FG$ occurs, while $DF$, given by
\begin{eqnarray}
DF:~~q_s =\xi_s^{-1}~,
\label{DFdef}
\end{eqnarray}
 plays the same role on the $q_z\xi_z^N \ll
\left(q_s\xi_s^N\right)^{\zeta_z}$
side.

 The locus $OF$, given by
\begin{eqnarray}
OF:~~q_z = {\xi_s\over\xi_z}q_s,\label{}\\
\end{eqnarray}
determines which of $B(T)q_z^2$ and $D(T)q_{\perp}^2$ dominates the denominators in
Eq. (\ref{CfunctionA}).

The locus $BC$, given by
\begin{eqnarray}
BC:~~q_z = \left(\xi_z^r\right)^{-1}~,
\label{BCdef}\\
\end{eqnarray}
is where we crossover between the random compression and the random field in Eq. (\ref{CfunctionA}),
while $AC$, given by
\begin{eqnarray}
AC:&&q_s = \left(\xi_s^r\right)^{-1},\label{}
\end{eqnarray}
plays the same role between the random tilt and the random field.


In these six regions, the $\vec{q}$-dependence of $\overline{\langle |u(\vec{q})|^2\rangle}$ is given by
\begin{widetext}
\begin{eqnarray}
 \overline{\langle |u(\vec{q})|^2\rangle}\propto\left\{
 \begin{array}{ll}
 \left(q_z\xi_z^N\right)^{{2\eta_K-\eta_t-6\over\zeta_z}}, &\mbox{region 1}\\
 \left(q_s\xi_s^N\right)^{2\eta_K-\eta_t-6}, &\mbox{region 2}\\
 \left(\xi_s^N\over\xi_s\right)^{{2\eta_K-\eta_t-6}}\left(q_z\xi_z\right)^{-2}, &\mbox{region 3}\\
 \left(\xi_s^N\over\xi_s\right)^{2\eta_K-\eta_t-6}\left(q_s\xi_s\right)^{-2}, &\mbox{region 4}\\
 \left(\xi_s^N\over\xi_s\right)^{1-{\eta_K\over 2}-\eta_B}\left(q_z\xi_z^N\right)^{-3}, &\mbox{region 5}\\
 \left(\xi_s^N\over\xi_s\right)^{\eta_K+{1\over 2}\eta_B-2}\left(q_s\xi_s^N\right)^{-3},&\mbox{region 6}
 \end{array}
 \right.
\end{eqnarray}
\end{widetext}
where, just to summarize everything in one place, the boundaries of the six regions are given by:
\begin{eqnarray}
FG:&&q_z\xi_z^N =
\left(q_{\perp}\xi_s^N\right)^{\zeta_z},\label{}\\
EF:&&q_z =\xi_z^{-1},\label{}\\
DF:&&q_s =\xi_s^{-1},\label{}\\
OF:&&q_z = {\xi_s\over\xi_z}q_s,\label{}\\
BC:&&q_z = \left(\xi_z^r\right)^{-1},\label{}\\
AC:&&q_s = \left(\xi_s^r\right)^{-1}.\label{}
\end{eqnarray}

This in turn implies for the light scattering:
\begin{widetext}
\begin{eqnarray}
 I(\vec{q})\propto\left\{
 \begin{array}{ll}
 q_s^2\left(q_z\xi_z^N\right)^{{2\eta_K-\eta_t-6\over\zeta_z}}, &\mbox{region 1}\\
q_s^{2\eta_K-\eta_t-4} \left(\xi_s^N\right)^{2\eta_K-\eta_t-6}, &\mbox{region 2}\\
q_s^2 \left(\xi_s^N\over\xi_s\right)^{{2\eta_K-\eta_t-6}}\left(q_z\xi_z\right)^{-2}, &\mbox{region 3}\\
 \left(\xi_s^N\over\xi_s\right)^{2\eta_K-\eta_t-6}\left(\xi_s\right)^{-2}, &\mbox{region 4}\\
q_s^2 \left(\xi_s^N\over\xi_s\right)^{1-{\eta_K\over 2}-\eta_B}\left(q_z\xi_z^N\right)^{-3}, &\mbox{region 5}\\
 \left(\xi_s^N\over\xi_s\right)^{\eta_K+{1\over 2}\eta_B-2}\left(\xi_s^N\right)^{-3}q_s^{-1},&\mbox{region 6}
 \end{array}
 \right.
\end{eqnarray}
\end{widetext}

\newpage
\newpage

The two characteristic lengths $\xi_z^r$ and $\xi_s^r$ are obtained by equating the random field piece ($\sqrt{D(T)B(T)\gamma(T)}$) of (\ref{CfunctionA}) to the random tilt and random compression pieces, in the appropriate regime of $\vec{q}$, and are given by
\begin{eqnarray}
\xi_s^r&=&\xi_s^N\left(\xi_s\over\xi_s^N\right)^{\Gamma_s},\label{Gamma_s}\\
\xi_z^r&=&\xi_z^N\left(\xi_s\over\xi_s^N\right)^{\Gamma_z},
 \end{eqnarray}
with
\begin{eqnarray}
 \Gamma_s&=&2-\eta_K+\eta_t+{1\over 2}\eta_B,\\
 \Gamma_z&=&3-{3\over 2}\eta_K+\eta_t.
\end{eqnarray}

Now imagine experiments in which one varies the light scattering wavevector $\vec{q}$ along the two different experimental loci we have indicated
in Fig. \ref{fig: lightsz}.


Along locus (1) we fix $q_{z,h}$ at 0 and vary $q_s$. We predict
\begin{widetext}
\begin{eqnarray}
 I(q_s)\propto\left\{
 \begin{array}{ll}
 \left(q_s\xi_s^N\right)^{2\eta_K-\eta_t-4}, &q_s\gg\xi_s^{-1},\\
 \left(\xi_s^N\over\xi_s\right)^{2\eta_K-\eta_t-4}, &\left(\xi_s^r\right)^{-1}\ll q_s\ll\xi_s^{-1},\\
 \left(\xi_s^N\over\xi_s\right)^{\eta_K+{1\over 2}\eta_B-2}\left(q_s\xi_s^N\right)^{-1}, &q_s\ll\left(\xi_s^r\right)^{-1}.
 \end{array}
 \right.
\end{eqnarray}
\end{widetext}
Thus, a  log-log plot of $I(q_s)$ verses $q_s$ will be a straight line with slope
$2\eta_K-\eta_t-4\approx-3.5$ for $q_s\gg\xi_s^{-1}$, where in the approximate equality we have used our $\epsilon$-expansion estimates (\ref{epsK}) and  (\ref{epsDt}) for $\eta_K$ and $\eta_t$. This straight line is then replaced by a horizontal section for
$\left(\xi_s^r\right)^{-1}\ll q_s\ll\xi_s^{-1}$, followed by another straight section with slope $-1$ for smaller $q_s\,$'s. We can read off $\xi_s$ and $\xi_s^r$ from
the place where the slope changes. Fitting $\xi_s^{-1}$ to $\left(T-T_{AC}\right)^{-\nu_s}$ determines $\nu_s$. Knowing $\xi_s$ and $\xi_s^r$  allows us to calculate $\Gamma_s$ from Eq. (\ref{Gamma_s}).

Along locus 2 we vary $q_z$ and keep $q_s$ fixed at a value in the range $0\ll q_s\ll\left(\xi_s^r\right)^{-1}$.
We find
\begin{widetext}
\begin{eqnarray}
 I(q_z)\propto\left\{
 \begin{array}{ll}
 \left(q_z\xi_z^N\right)^{{2\eta_K-\eta_t-6\over\zeta_z}}, &q_z\gg\xi_z^{-1},\\
 \left(\xi_s^N\over\xi_s\right)^{{2\eta_K-\eta_t-6}}\left(q_z\xi_z\right)^{-2}, &\left(\xi_z^r\right)^{-1}\ll q_z\ll\xi_z^{-1},\\
 \left(\xi_s^N\over\xi_s\right)^{1-{\eta_K\over 2}-\eta_B}\left(q_z\xi_z^N\right)^{-3}, &\left(\xi_s\over\xi_z\right)q_s\ll q_z\ll\left(\xi_z^r\right)^{-1},\\
 \left(\xi_s^N\over\xi_s\right)^{\eta_K+{1\over 2}\eta_B-2}\left(q_s\xi_s^N\right)^{-3},&q_z\ll\left(\xi_s\over\xi_z\right)q_s.
 \label{Ilocus2}
 \end{array}
 \right.
\end{eqnarray}
\end{widetext}
Using an approach  similar to that just described for locus 1, we can obtain $(2\eta_K-\eta_t-4)/\zeta_z$, $\xi_z$, $\xi_z^r$, and hence $\nu_z$, and $\Gamma_z$. These results combined with those from locus 1 allow us to find $\eta_{B,K,t}$. Then plugging $\nu_s$ and $\eta_{B,K,t}$ into Eqs. (\ref{ExactNu_h}, \ref{ExactEta_c}) gives $\eta_c$ and $\nu_h$, respectively.

Finally, since $\eta_{B,K,t}$, $\nu_{s,z}$, $\beta$, and $\alpha$ can all be measured experimentally, the exact scaling relations Eqs. (\ref{ExactBeta}, \ref{ExactAlpha}, \ref{ExactNu_z}) can be examined.

In the $C$ phase the relation between the light scattering intensity and the fluctuations
is given by a formula similar to (\ref{LightScattering}):
\begin{eqnarray}
I(\vec{q})\sim\sum_{ij}E'_{ij}C'_{ij}(\vec{q})\label{}
\end{eqnarray}
where
\begin{eqnarray}
 C'_{ij}(\vec{q})\equiv L_{ij}(\hat{q})q_{\perp}^2\overline{\langle |u'(\vec{q})|^2\rangle}.
\end{eqnarray}
The coefficients $E'_{ij}$ are independent of $\vec{q}$. However, their value is affected
by the tilt angle of the layer normals, and therefore has a weak dependence on the temperature.
Since we have argued in Sec. \ref{Sec: phases} that $\overline{\langle |u'(\vec{q})|^2\rangle}$
has the same expression as $\overline{\langle |u(\vec{q})|^2\rangle}$, we conclude that the $C$ phase
has qualitatively the same light scattering behavior as the $A$ phase.


\section{\label{Sec: Stability} Stability of the smectic phase at the critical point}
All of our discussion so far has assumed that the smectic state remains stable right up to the AC critical point. There are two conceivable mechanisms by which this assumption might fail: first, the orientational order of the smectic layers could be destroyed; and second, the integrity of the smectic layers themselves could fail via the proliferation of dislocations in the smectic layers.

In this section, we argue that neither of these happens, and that the smectic state is stable right up to, and through, the AC transition.

Consider first  the orientational order of the system.  In deriving our model we have implicitly assumed that the both the real space fluctuations $\overline{\langle|\delta\vec{n}(\vec{r})|^2\rangle}$ of the nematic directors  $\hat{n}(\vec{r})$, and those $\overline{\langle|\delta\vec{N}(\vec{r})|^2\rangle}$ of the normals  $\hat{N}(\vec{r})$ to the smectic layers, are small, so that the expansions $\hat{n}=\hat{z}+\delta \vec{n}$ and $\hat{N}=\hat{z}+\delta \vec{N}$ are justified. Thus, for our theory to be valid, long-ranged orientational order must exist.
To show that it does, even at the AC critical point, at least for sufficiently small disorder, we need to calculate $\overline{\langle|\delta\vec{n}(\vec{r})|^2\rangle}$ and $\overline{\langle|\delta\vec{N}(\vec{r})|^2\rangle}$.

We can obtain both of these quantities by
calculating $\overline{\langle|\vec{\nabla} _\perp u|^2\rangle}$, since, as shown in section \ref{Sec: Model},  fluctuations of both $\delta\vec{N}(\vec{r})$ and $\delta\vec{n}(\vec{r})$ away from $\vec{\nabla} _\perp u$ are massive. Hence,  up to small corrections,
 $\overline{\langle|\delta\vec{n}(\vec{r})|^2\rangle}= \overline{\langle|\delta\vec{N}(\vec{r})|^2\rangle}=\overline{\langle|\vec{\nabla} _\perp u|^2\rangle}$.


 The most problematical point is right at the AC transition, since there the fluctuations are  biggest. Hence, if
$\overline{\langle|\delta\vec{n}(\vec{r})|^2\rangle}$ and $\overline{\langle|\delta\vec{N}(\vec{r})|^2\rangle}$ are finite at this point, they'll be finite everywhere.

We will now show that they are, indeed, finite right at the AC transition.

Because of the anisotropic scaling,  $\overline{\langle(\partial_s u(\vec{r})^2\rangle}$
 is the dominant part of $\overline{\langle|\vec{\nabla}_\perp u|^2\rangle}$. Using our earlier equation (\ref{u_qc}) for
$\overline{\langle| u(\vec{q})|^2\rangle}$, we obtain:
 \begin{eqnarray}
  \overline{\langle(\partial_s u)^2\rangle}&=&{1\over (2\pi)^3}\int d^3q~{\Delta_t(\vec{q}) q_s^2\over \left[B(\vec{q})q_z^2+\gamma q_h^2+K(\vec{q})q_s^4\right]^2} \nonumber\\
                    &\propto& L^{\Gamma'}
 \end{eqnarray}
with
\begin{eqnarray}
\Gamma'=-1+\eta_t+\eta_K+{\eta_B\over 2},
\end{eqnarray}
where $L$ is the infrared cutoff in the $s$ directions. Thus, $\Gamma'<0$
is required for the convergence of $\overline{\langle|\vec{\nabla}u|^2\rangle}$ and hence the convergence of $\overline{\langle(\delta n_i)^2\rangle}$. Clearly, in $d=3$ our numerical estimate to  first order in $\epsilon$ does satisfy this condition.
This shows that at the AC transition point point, the smectic is stable
against orientational fluctuations.

In previous sections we have shown that right at the critical point the long-ranged translational order is destroyed
in the presence of the quenched disorder. Thus one might reasonably question what is the difference between the smectic
and the isotropic (nematic) phases, and even worry whether the smectic phase survives at the $AC$ critical point at all. However, there are examples
in which the low-temperature and the high-temperature phases are not distinguished by the order of the system, but rather,
by whether the topological
defects are bound or unbound. The most famous one is the Kosterlitz-Thouless transition in the $d=2$ XY
model \cite{KT}. Another famous example is the ``Bragg glass'' in pinned superconducting flux lines \cite{TN, TG, Huse, Fisher}.
We will now show that the same thing happens in our problem.
Since similar calculations have been given in Refs \cite{RT, Karl} in great detail, and ours are virtually the same,
our description will be very brief.

The starting point of the theory is the tilt only harmonic Hamiltonian at the critical point:
\begin{eqnarray}
H &=& \int d^dr  \left[{ K \over 2}(\partial_s^2u)^2 + {B
\over
2}(\partial_zu)^2\right.\nonumber\\
&&\left.+{\gamma\over 2}(\partial_hu)^2+\vec{h}\cdot\vec{\nabla} u\right].
\label{Squadratic}
\end{eqnarray}
where the random field disorder is not included. It can be justified
that the random field disorder is irrelevant when
dislocations are included.

When the smectic contains  dislocations, the displacement field $u$ is
no longer single valued. Mathematically this can be represented as
\begin{eqnarray}
\vec{\nabla}\times\vec{\nabla} u=\vec{m} \label{defineM}
\end{eqnarray}
with
\begin{eqnarray}
\vec{m}(\vec{r})=\sum_i\int{aM_it(s_i)\delta^3\left(\vec{r}-
\vec{r}_i(s_i)\right)ds_i}\, ,
\end{eqnarray}
where $s_i$ parameterizes the path of  the $i$'th
dislocation loop, $M_i$ is an integer giving the charge of that
loop, $t(s_i)$ is the local tangent of the loop, and $\vec{r}_i$
is a position of the point on the loop parameterized by $s_i$ the loop. Furthermore, equation
(\ref{defineM}) implies
\begin{eqnarray}
 \vec{\nabla}\cdot\vec{M}=0,
\end{eqnarray}
which means that dislocation lines can not end in the bulk of the
sample.

To obtain a dislocation Hamiltonian, we need to trace over field
$u$ which is constrained by equation (\ref{defineM}). This can be
done in the following standard way. We separate the
field $\vec{v} = \vec{\nabla}u$ into
\begin{eqnarray}
 \vec{v} =  \vec{v}_d + \delta\vec{v}
 \label{decomposition},
\end{eqnarray}
where $\vec{v}_p$ minimizes Hamiltonian (\ref{Squadratic}) for a
given dislocation configuration $\vec{m}(\vec{r})$,
$\delta\vec{v}$ can be viewed as the fluctuation from the ground
state. Inserting (\ref{decomposition}) into Hamiltonian
(\ref{Squadratic}), we find that $\vec{v}_p$ and $\delta\vec{v}$
are decoupled due to the construction that $\vec{v}_d$
minimizes Hamiltonian (\ref{Squadratic}). Thus we obtain the
effective model for $\vec{m}(\vec{r})$:
\begin{eqnarray}
H_d&=&\sum_{\vec{q}}\left[{1\over 2B\Gamma_q}\left(K\gamma q_s^2|m_z(\vec{q})|^2+BK q_s^2|m_h(\vec{q})|^2\right.\right.\nonumber\\
   &&\left.\left.+B\gamma |m_s(\vec{q})|^2\right)+\vec{m}(\vec{q})\cdot\vec{a}(\vec{q})\right],
 \label{defectH}
\end{eqnarray}
where $\Gamma_q=q_z^2+(\gamma/B)q_h^2+\lambda_z^2q_s^4$, and
\begin{eqnarray}
\vec{a}=i\left[{\vec{q}\times\vec{h}\over
q^2}+{\left[\left(Bq_z\hat{z}+\gamma q_h\hat{h}+Kq_s^3\hat{s}\right)\times\vec{q}\right]\vec{q}\cdot\vec{h}\over B\Gamma_qq^2}
\right].\nonumber\\
 \label{fielda}
\end{eqnarray}

By putting the model on a a simple cubic lattice, the partition
function can be written as
\begin{eqnarray}
 Z=\sum_{\vec{m}{\vec{r}}}e^{-S[\vec{m}]}\, ,
 \label{Partition1}
\end{eqnarray}
where
\begin{eqnarray}
 \vec{m}(\vec{r}) &=& {a\over d^2}[n_x(\vec{r}), n_y(\vec{r}),
 n_z(\vec{r})],
 \label{discretem}
 \\
 S[\vec{m}] &=&{1\over T}\left[H_d[\vec{m}]+{E_cd^4\over
 a^2}\sum_{\vec{r}}|\vec{m}(\vec{r})|^2\right], \nonumber \\
\end{eqnarray}
where the $n_i$'s are integers, $d$ is the cubic lattice
constants used in the discretization, and ${E_cd^4\over
 a^2}\sum_{\vec{r}}|\vec{m}(\vec{r})|^2$ is the core energy term
 coming from the core of the defect line that is not accurately
 treated by the continuum elastic theory.

To cope with the constraint $\vec{\nabla}\cdot\vec{m} = 0$, we
introduce an auxiliary field $\theta(\vec{r})$, rewriting the
partition function equation (\ref{Partition1}) as
\begin{eqnarray}
 Z = \prod_{\vec{r}}\int d\theta(\vec{r}) \sum_{\vec{m}(\vec{r})}
 e^{-S[\vec{m}]+i\sum_{\vec{r}}\theta(\vec{r})\vec{\nabla}\cdot
 \vec{m}(\vec{r})d^2/a},~~~~~~
\end{eqnarray}
where the sum over $\vec{m}(\vec{r})$ is now unconstrained.

Then we introduce a dummy vector field $\vec{A}$ to mediate the
long-range interaction between defect loops in the Hamiltonian
(\ref{defectH}). This is accomplished by rewriting the partition
function as
\begin{eqnarray}
 Z = \prod_{\vec{r}}\int d\theta(\vec{r})d\vec{A}(\vec{r})
 \sum_{\vec{m}(\vec{r})}
 e^{-S[\vec{m}, \theta, \vec{a}]}\delta(\vec{\nabla}\cdot\vec{A})
 \delta(A_z)\nonumber\\
 \label{Partition2}
\end{eqnarray}
with
\begin{eqnarray}
 S&=&{1\over T}\sum_{\vec{r}}\left[\vec{m}\cdot \left(-i{Td^2\over
 a}\vec{\nabla}\theta(\vec{r})+d^3[i\vec{A}(\vec{r})+\vec{a}(\vec{r})]\right)
 \right.\nonumber\\
 &~&\left.+E_c{d^4\over a^2}|\vec{m}|^2\right]+{1\over 2T}
 \sum_{\vec{q}}\left({B\Gamma_q\over K\gamma q_s^2}|A_z(\vec{q})|^2+\right.\nonumber\\
 &~&\left.{\Gamma_q\over Kq_s^2}|A_h(\vec{q})|^2+{\Gamma_q\over\gamma}|A_s(\vec{q})|^2\right).
\end{eqnarray}

The sum over $\vec{m}(\vec{r})$ is just  the
``periodic Gaussian'' introduced by Villain \cite{Villain}. Then the
partition function equation (\ref{Partition2}) can be rewritten as
\begin{eqnarray}
 Z &=& \prod_{\vec{r}}\int d\theta(\vec{r})d\vec{A}(\vec{r})
 \ \ \delta(\vec{\nabla}\cdot\vec{A})\delta(A_z)
 \nonumber\\
 &~&\times \exp\left[-\sum_{\vec{r}i}V_p
 \left[\theta(\vec{r}+\hat{x}_i)-\theta(\vec{r})
 -{ad\over T}[A_i(\vec{r})\right.\right.\nonumber\\
 &~&\left.\left.-ia_i(\vec{r})]\right]
 -{1\over 2T}\sum_{\vec{q}}\left({B\Gamma_q\over K\gamma q_s^2}|A_z(\vec{q})|^2+\right.\right.\nonumber\\
 &~&\left.\left.{\Gamma_q\over Kq_s^2}|A_h(\vec{q})|^2+{\Gamma_q\over\gamma}|A_s(\vec{q})|^2\right)\right],
 \label{dualmodel}
\end{eqnarray}
where the $2\pi$-period Villain potential $V_p(x)$ is defined as
\begin{eqnarray}
 e^{-V_p(x)} \equiv \sum_{-\infty}^{\infty}
 e^{-n^2E_c/T + ixn}.
\end{eqnarray}
Since $V_p(x)$ has sharper minima for {\it smaller} $E_c/T$,
which corresponds to higher temperature, raising the temperature
in the original model is equivalent to lowering temperature in the
dual model equation (\ref{dualmodel}).

Standard universality class arguments imply that the model
equation (\ref{dualmodel}) has the same universality class as the
``soft spin'', or Landau-Ginsburg-Wilson, with the {\it complex}
``action''
\begin{eqnarray}
S_r&=&\sum_{r,\alpha}\left[{c\over 2}\left(\vec{\nabla}+{ad\over
T}(i\vec{A}_{\alpha}+\vec{a})\right)\psi_{\alpha}^*\cdot
\left(\vec{\nabla}-{ad\over T}(i\vec{A}_{\alpha}+\right.\right.\nonumber\\
&~&\left.\left.\vec{a}\right)\psi_{\alpha}+t_d|\psi_{\alpha}|^2+u_d|\psi_{\alpha}|^4\right]
+\sum_{\vec{q}}\left({B\Gamma_q\over K\gamma q_s^2}|A_z(\vec{q})|^2\right.\nonumber\\
 &~&\left.+{\Gamma_q\over Kq_s^2}|A_h(\vec{q})|^2+{\Gamma_q\over\gamma}|A_s(\vec{q})|^2\right)\, ,\nonumber\\
 \label{softspin}
\end{eqnarray}
where $\psi$ is a complex order parameter whose phase is
$\theta(\vec{r})$. Because of the duality transformation's
inversion of the temperature axis, the reduced temperature $t_d$
is a monotonically decreasing function of the temperature $T$ (of
the original dislocation loop model), which vanishes at the
mean-field transition temperature of the dislocation model
({\ref{dualmodel}). Disorder is included in the model
(\ref{softspin}) through $\vec{a}(\vec{r})$, which is related to
the random tilt field $\vec{h}(\vec{r})$ by equation
(\ref{fielda}).

Because of the duality inversion of the temperature axis,
the ordered phase of the dual model (\ref{softspin})
corresponds to the disordered phase (dislocation loops unbound) of
the original dislocation model.

An complete analysis of the dislocation loop unbinding transition
described by the model (\ref{softspin}) is beyond the scope of this
paper. The goal here is to know if a dislocation bound state ever
exists near the critical point of the $AC$
transition. Let us check the one-loop
graphical correction to the dual temperature:
\begin{eqnarray}
t_R=t_0-{ca^2d^2\over T^2}\int{{d^3q\over
(2\pi)^3}{\Delta_t\left(B^2q_z^2+\gamma^2 q_h^2\right)\over B^2\Gamma^2_q}}.~~~
 \label{Correctiont}
\end{eqnarray}
If $K$ and $\Delta_t$ are treated as constant it is easy to show that the integral diverges.
This result, if true, would imply that the smectic phase becomes unstable against dislocation unbinding sufficiently close to the putative AC transition.

However, this conclusion only holds within the {\it harmonic} approximation.
In sections \ref{Sec: RG} and \ref{Sec: Anomalous Elasticity} we have shown that  anharmonic
effects are important. A crude way
to include their effects is to replace the constant $K$ and $\Delta$ in the harmonic calculation above with the anomalous, wavevector-dependent $K(\vec{q})$ and
$\Delta_t(\vec{q})$ found earlier in section \ref{Sec: Anomalous Elasticity}. Implementing this in the integral
and requiring it to converge, we obtain a
restriction on $\eta_K$ and $\eta_t$:
\begin{eqnarray}
1-\eta_t-{\eta_B\over 2}>0\, .
\end{eqnarray}
This condition is satisfied by the values of
$\eta_B$ and $\eta_t$ obtained by the $\epsilon$-expansion to $O(\epsilon)$
in the physical dimension $d=3$. This implies that the
the second-order smectic $A$-$C$ phase transition is stable against dislocation
unbinding.

\section{Conclusion}
We have developed a theory of the smectic-A to smectic-C phase transition
in a biaxial disordered environment. Using an $\epsilon={7\over 2}-d$ expansion,  we have shown that
the phase transition can be second order. The
critical exponents were  calculated to  first order in $\epsilon$. We also calculated the smectic correlation
functions near the critical point, which lead to exotic X-ray and light scattering patterns. All these predictions should be
testable provided that the biaxial disordered environment can be realized in experiments.

\section{Acknowledgements}
L.C. acknowledges support by NSF of China (under Grant
No. 11004241) and the Fundamental Research Funds for the
Central Universities (under Grant No. 2010LKWL09). J. T. thanks  the Initiative for the Theoretical Sciences at The Graduate Center of the City University of New York,  the Lorentz Center of Leiden University, and the Max-Planck-Institut fur Physik Komplexer Systeme, Dresden, for their hospitality while this work was underway.

\appendix
\section{\label{Sec: SoftContinuation}}
There are two ways of continuing our model Eq. (\ref{CompleteH}) from three dimensions to higher spatial dimensions. In the main text
we have focused on the ``hard" continuation, in which we keep the $h$-space  $d-2$ dimensional
and the $s$-space  one dimensional. In this appendix we briefly describe the calculation which uses the soft continuation, in
which we keep the $h$-space one dimensional and the $s$-space $d-2$ dimensional.

The RG procedure is essentially the same as that used in the hard continuation. After integrating out the freedom
of the fast varying part of $u(\vec{r})$ and rescaling, we obtain the following RG flow equations:
\begin{eqnarray}
{d\gamma\over d\ell}&=&\left[d-\omega_h+\omega_z+2\chi-2\right]\gamma,\\
{dB\over d\ell}&=&\left[d+\omega_h-\omega_z+2\chi-2-{1\over
4}\tilde{g}_1\right]B,\\
{dK\over d\ell}&=&\left[d+\omega_h+\omega_z+2\chi-6+{1\over
8}\tilde{g}_1\right]K,\label{}\\
{dg\over d\ell}&=&\left[d+\omega_h+3\chi-4+{1\over 4}\tilde{g}_1-{1\over 2}\tilde{g}_2\right]g,\nonumber\\
\label{}\\
{dw\over d\ell}&=&\left[d+\omega_h+\omega_z+4\chi-6-{3\over
8}{\tilde{g}_1^2\over \tilde{g}_2}\right]w\nonumber\\&&+\left[{5\over
4}\tilde{g}_1-{9\over 8}\tilde{g}_2\right]w,\\
{d\Delta_t\over d\ell}&=&\left[d+\omega_h+\omega_z+2\chi-4+{1\over 24}\tilde{g}_1\right]\Delta_t,\\
{dD\over d\ell}&=&\left[d+\omega_h+\omega_z+2\chi-4+{3\over
8}(\tilde{g}_1-\tilde{g}_2)\right]D\nonumber\\&&+{3\over 8}K(\tilde{g}_2-\tilde{g}_1),\label{SoftD}
\label{}
\end{eqnarray}
where $\tilde{g}_1$ and $\tilde{g}_2$ are two dimensionless couplings defined by:
\begin{eqnarray}
\tilde{g}_1&\equiv& C_{d-2}(g/B)^2\Delta_t
(B/\gamma)^{1/2}K^{-2}\Lambda^{d-4}~, ~~~~~~~~\\
\tilde{g}_2&\equiv& C_{d-2}(w/B)\Delta_t
(B/\gamma)^{1/2}K^{-2}\Lambda^{d-4}~.\nonumber\\
\end{eqnarray}
These couplings $\tilde{g}_1$ and $\tilde{g}_2$ flow according to
\begin{eqnarray}
 {d\tilde{g}_1\over d\ell}&=&\tilde{\epsilon}+{2\over 3}\tilde{g}_1-\tilde{g}_2,\label{Softg1}\\
 {d\tilde{g}_2\over d\ell}&=&\tilde{\epsilon}-{3\over 8}{\tilde{g}_1^2\over g_2}-{9\over 8}\tilde{g}_2+{7\over 6}\tilde{g}_1~,\label{Softg2}
\end{eqnarray}
where $\tilde{\epsilon}=4-d$.

Once again, it is convenient but not necessary to make a special choice of $\chi$, $\omega_h$, and $\omega_z$ such that $B$, $K$, and $\gamma$ are
fixed at their bare values. This requirement can only be fulfilled by
\begin{eqnarray}
 \omega_h&=&2-{1\over 16}\tilde{g}_1,\\
 \omega_z&=&2-{3\over 16}\tilde{g}_1,\\
 \chi&=&{2-d\over 2}+{1\over 16}\tilde{g}_1.
\end{eqnarray}
This choice turns flow Eq. (\ref{SoftD}) into
\begin{eqnarray}
 {dD\over d\ell}&=&\left[2-{1\over 8}\tilde{g}_1+{3\over
8}(\tilde{g}_1-\tilde{g}_2)\right]D+{3\over 8}K(\tilde{g}_2-\tilde{g}_1).\nonumber\\
\end{eqnarray}

Flow Eqs. (\ref{SoftD},\ref{Softg1},\ref{Softg2}) have three fixed points in total. One of them is the trivial Gaussian fixed point
$g_1^*=g_2^*=0$, $D=0$. Another one was also found previously for the  $m=1$ Bragg Glass problem\cite{Karl}, which preserves the symmetry that the Hamiltonian is invariant under any rigid rotation of the liquid crystal about the $h$ axis. This fixed point is:
\begin{eqnarray}
 \tilde{g}_1^*&=&\tilde{g}_2^*=3\tilde{\epsilon}+O(\tilde{\epsilon}^2),\\
 D^*&=&0,
\end{eqnarray}

Both of these fixed points  are unstable in two or more directions.

The third fixed point is:
\begin{eqnarray}
 \tilde{g}_1^*&=&{3\over 7}\tilde{\epsilon}+O(\tilde{\epsilon}^2),\\
 \tilde{g}_2^*&=&{9\over 7}\tilde{\epsilon}+O(\tilde{\epsilon}^2),\\
 D^*&=&-{9K\over 56}\tilde{\epsilon}+O(\tilde{\epsilon}^2),
\end{eqnarray}
which is stable in two directions, and unstable in a third. Therefore, this fixed point  controls the
critical behavior of the phase transition.

With this fixed point in hand,  the anomalous elasticity and the critical exponents can be calculated by following the steps
described in Secs. \ref{Sec: Anomalous Elasticity} and \ref{Sec: Critical Behavior}. Here we only give the
results:
\begin{eqnarray}
 \tilde{\eta}_B&=&{3\over 28}\tilde{\epsilon}+O(\tilde{\epsilon}^2),\\
 \tilde{\eta}_K&=&{3\over 56}\tilde{\epsilon}+O(\tilde{\epsilon}^2),\\
 \tilde{\eta}_t&=&{1\over 56}\tilde{\epsilon}+O(\tilde{\epsilon}^2),\\
 \tilde{\nu}_s&=&{1\over 2}+{3\over 32}\tilde{\epsilon}+O(\tilde{\epsilon}^2),\\
 \tilde{\nu}_h&=&1+{39\over 224}\tilde{\epsilon}+O(\tilde{\epsilon}^2),\\
 \tilde{\nu}_z&=&1+{33\over 224}\tilde{\epsilon}+O(\tilde{\epsilon}^2),\\
 \tilde{\beta}
 &=&{1\over 2}-{9\over 32}\tilde{\epsilon}+O(\tilde{\epsilon}^2)\\
 \tilde{\alpha}
 &=&{9\over 56}\tilde{\epsilon}+O(\tilde{\epsilon}^2).
\end{eqnarray}

Now we compare the numerical results obtained from the two continuations. In $d=3$, to  first order in $\epsilon$ and $\tilde{\epsilon}$ we
find, for the hard continuation discussed earlier,
\begin{eqnarray}
 \eta_B&=&0.2222,\\
 \eta_K&=&0.2963,\\
 \eta_t&=&0.07407,\\
 \nu_s&=&0.6296,\\
 \nu_h&=&1.1852\\
 \nu_z&=&1.1296,\\
 \beta&=&0.3889,\\
 \alpha&=&0.1111.
\end{eqnarray}
while for the soft continuation treated in this appendix,
\begin{eqnarray}
 \tilde{\eta}_B&=&0.1071,\\
 \tilde{\eta}_K&=&0.05357,\\
 \tilde{\eta}_t&=&0.01786,\\
 \tilde{\nu}_s&=&0.5938,\\
 \tilde{\nu}_h&=&1.1741,\\
 \tilde{\nu}_z&=&1.1473,\\
 \tilde{\beta}&=&0.2188,\\
 \tilde{\alpha}&=&0.1607,
\end{eqnarray}
Unfortunately, the two sets of results are not in good agreement.
We trust the hard continuation more, since $\epsilon$ is smaller ($\epsilon={1\over 2}$) for that configuration than the expansion parameter $\tilde{\epsilon}=1$ for the soft continuation.

\section{\label{sec: AnomalousElasticityIrrelevant}Wavevector-dependences of Irrelevant Disorder Variances}
In our RG calculations in section \ref{Sec: RG}, we did not
include in the Hamiltonian (\ref{SimpleH}) the random compression disorder
\begin{eqnarray}
-{\Delta_c\over 2T}\int d^dr \sum^n_{\alpha, \beta = 1}
\partial_zu_{\alpha} \cdot \partial_zu_{\beta}\, ,
\end{eqnarray}
because simple power counting shows that this term is irrelevant.
However, while it has no effect on the critical exponents, the
disorder invariance $\Delta_c$ itself develops strong power-law
dependence on wavevector at long wavelengths. We need to know
this power-law exponent to calculate the fluctuations
caused by the random compression, which have a significant effect on the
light scattering intensity.

\begin{figure}
 \includegraphics[width=0.4\textwidth]{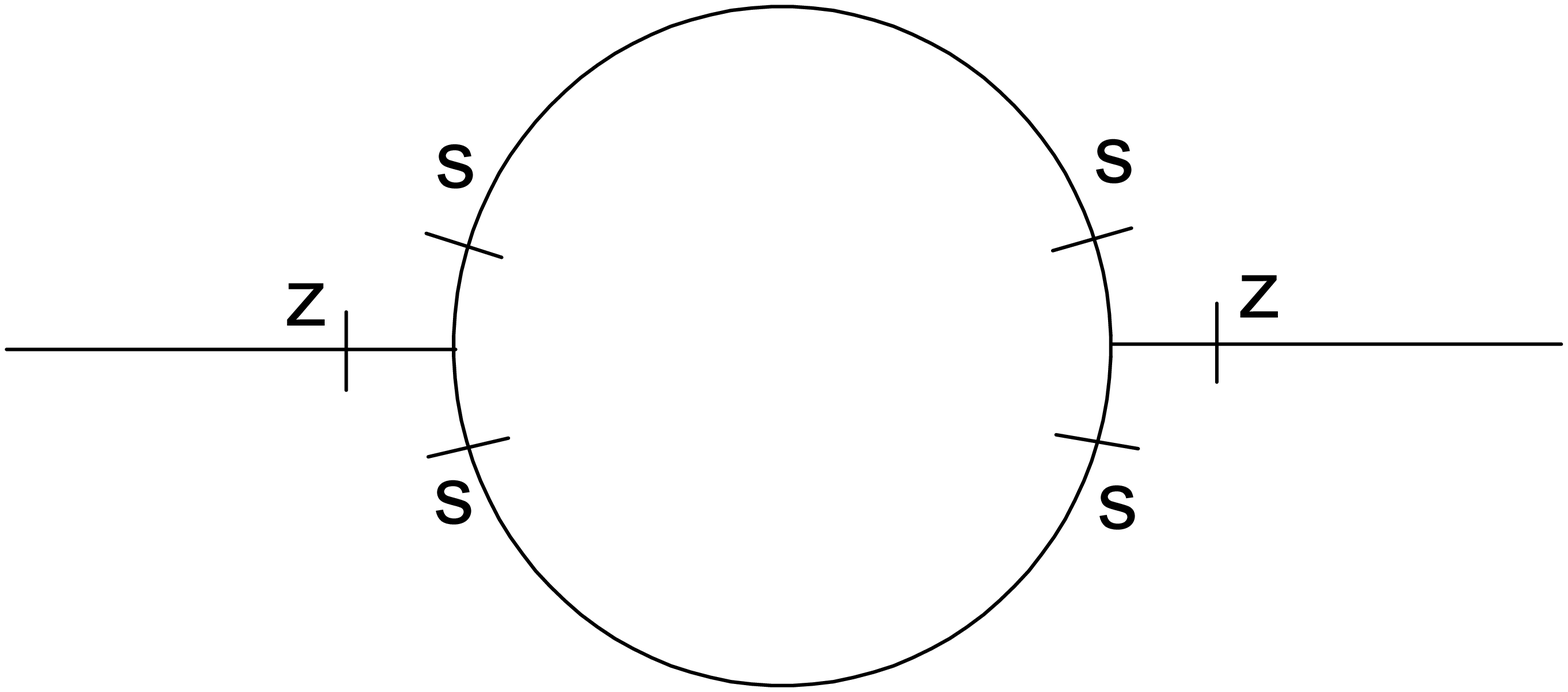}
 \caption{\label{fig: DeltaC}Graphical correction to $\Delta_c$.}
\end{figure}

The most important one-loop graphical correction to $\Delta_c$ comes
from the Feynman diagram in Fig. \ref{fig: DeltaC}, from which
we obtain
\begin{eqnarray}
\delta\Delta_c  &=& {g^2 \over 2} \int ^>_p \Delta_t^2
p^8_sG\left(\vec{p}\right)^4\nonumber \\
&=& {g^2\Delta_t^2\over 2} \int^{\infty}_{-\infty} {dp_z \over 2\pi}
\int^{\infty}_{-\infty} {d^{d-2}p_h \over (2\pi)^{d-2}}\int ^> {d p_s \over  (2\pi)}p^8_sG\left(\vec{p}\right)^4\nonumber\\
&=& {11\over 512\sqrt{2}}{\Delta_t B g_1\over K \Lambda^2}\label{}.
\end{eqnarray}
Thus, using the length  and field rescalings we used in section
\ref{Sec: RG}, the RG flow equation of $\Delta_c$ to one-loop order is
given by
%
\begin{eqnarray}
{d\Delta_c(\ell)  \over  d \ell} = \left[2\chi+\left(d-2\right)\omega_h - \omega_z + 1+{11\over 512\sqrt{2}}g_3\right]\Delta_c\nonumber, \\
\label{Delta_c}
\end{eqnarray}
where the dimensionless coupling $g_3$ is defined as
\begin{eqnarray}
g_3\equiv {B\Delta_t\over K\Delta_c\Lambda^2} g_1\, .\label{g3}
\end{eqnarray}
Combining this RG flow equation with the flow Eqs.
(\ref{Bflow}, \ref{Kflow}, \ref{Delta_t}, \ref{g1flow}), we get
\begin{eqnarray}
 {d g_3\over d \ell} = \left( 2 - {8\over 9}\epsilon -
 {11\over 512\sqrt{2}}g_3 \right)g_3 \label{},
\end{eqnarray}
which flows to a stable nonzero fixed point
\begin{eqnarray}
g_3^* = {\sqrt{2}\over 99}\left(9216-4096\epsilon\right).
\end{eqnarray}

Now the wavevector dependence of $\Delta_c$
can be calculated by using trajectory integral matching on the
correlation function $\overline{\langle u(-\vec{q})\rangle\langle
u(\vec{q})\rangle}=(\Delta_t q_s^2+\Delta_c q_z^2
)G(\vec{q})^2$, Performing a calculation which is essentially the same as those
in section \ref{Sec: Anomalous Elasticity}, we find
\begin{eqnarray}
\Delta_c(\vec{q})&=&\Delta_c\left(\xi_s^Nq_s\right)^{-\eta_c}f_c\left(X, Y\right)~,
 \label{AnomalousDelta_c}
\end{eqnarray}
with the anomalous exponent
\begin{eqnarray}
 \eta_c=2-{8\over 9}\epsilon+O(\epsilon^2)\, .
 \label{etac}
\end{eqnarray}
Here $f_c(X,Y)$ is another crossover function of the scaling variables $X$ and $Y$ given by (\ref{Xdef}) and (\ref{Ydef}), respectively.
Note that $\eta_c$ is nonzero even to  {\it zeroth} order in $\epsilon$,
which is quite common for irrelevant variables.

We can also obtain Eq. (\ref{etac}) through an {\it exact}
scaling relation between $\eta_c$ and other anomalous exponents.
This scaling relation is implied by the fact that $g_3$ flows to a non-zero
stable fixed point. For large enough $\ell$, $g_3$ reaches a
fixed point and thus
\begin{eqnarray}
{d \ln {g_3}\over d \ell}=0\, .
\end{eqnarray}
This equation, after rewriting $g_3$ in terms of other parameters using its definition
(\ref{g3}), leads to
\begin{eqnarray}
 {d \ln {B}\over d \ell}+{d \ln {\Delta_t}\over d \ell}+{d \ln {g_1}\over d \ell}
 -{d \ln {K}\over d \ell}-{d \ln {\Delta_c}\over d \ell}=0\,
 .\nonumber\\
\end{eqnarray}
Then we replace the graphical part in flow Eqs. (\ref{Bflow}, \ref{Kflow}, \ref{Delta_t},\ref{Delta_c})
by $-\eta_B$, $\eta_{K,t,c}$, respectively, and plug them and Eq. (\ref{g1flow}) into the above equation.
The rescaling factors $\chi$ and $\omega$ vanish, and we are left with
\begin{eqnarray}
\eta_c = 2 + \eta_t - \eta_B - \eta_K.  \label{Scalingeta_c}
\end{eqnarray}
Using the already known  values of $\eta_t$, $\eta_B$, and $\eta_K$ to $O(\epsilon)$, using (\ref{Scalingeta_c}) also leads to the epsilon expansion for $\eta_c$; reassuringly, this expansion is exactly (\ref{etac}).


\begin{thebibliography}{99}
\bibitem{brittle}Since aerogel tends to be rather brittle, it may be more practical to prepare aerogel with the desired type of anisotropy by growing it anisotropically, rather than by stretching and compressing an already formed isotropic aerogel. All that matters is that the final aerogel state be the same as one constructed by a gedanken experiment in which an originally isotropic aerogel is stretched and compressed in the manner we described here. This has already  been achieved experimentally for a liquid crystal in aerogel system; see, e.g., the last of reference \cite{LC}.
\bibitem{eta}While it is not usually interpreted this way, the existence of a non-zero critical exponent $\eta$ at most conventional ferromagnetic phase transitions can be interpreted as a signature of anomalous elasticity in the magnetic gradient energy elastic constant $c$, which enters the Hamiltonian as $\int d^dr~ c~ |\vec{\nabla}\vec{M} |^2$~~.
\bibitem{clean smectic}This is in contrast to clean (i.e., disorder-free) smectics, which also exhibit power law decay of translational correlations, due in that case to thermal fluctuations, but with a non-universal (and, indeed, temperature-dependent) exponent. Furthermore, the clean smectic peaks exhibit anisotropic scaling, while the RFXY peaks we describe here do not. See references \cite{Caille}.
\bibitem{easyRFXY} One easy way to see this is to imagine that the random field couples to the magnetization via a conventional external field-magnetization coupling Hamiltonian  $H_{field} = -\int d^dr [\vec{h}(\vec{r}) \cdot \vec{M}(\vec{r})]$  , with the $\vec{h}(\vec{r})$'s all of the same magnitude $|\vec{h}(\vec{r})|=h$, but oriented completely randomly. This then leads to a model precisely of the form (\ref{RFXY2}), with $V^{XY}_p(\theta-\phi_{XY}(\vec{r}))=-h \cos(\theta-\phi_{XY}(\vec{r}))$, and $\phi_{XY}(\vec{r})$ just the angle between the random field at point $\vec{r}$ and some refrence direction.
\bibitem{O(1) comment}One could more generally take $U(y)=Cy^4$, where $C=O(1)$; we can, without loss of generality, absorb this $O(1)$ constant into that in our definition (\ref{k def}) of $k(\vec{r})$, and we choose to do so.
\bibitem{saddle} Equation [\ref{saddle cond quartic}] also has a solution $x=-\left({1\over 4k^2}\right)^{1\over 3}i$; we ignore this because it proves to not be a saddle point.
\end{thebibliography}
\end{document}